\documentclass[journal]{IEEEtran}
\usepackage{amsmath,amsfonts}
\usepackage{algorithmic}
\usepackage{algorithm}
\usepackage{array}
\usepackage[caption=false,font=normalsize,labelfont=sf,textfont=sf]{subfig}
\usepackage{textcomp}
\usepackage{stfloats}
\usepackage{url}
\usepackage{verbatim}
\usepackage{graphicx}
\usepackage{cite}
\usepackage{bm}
\usepackage{amssymb}
\usepackage{multirow}     
\usepackage{diagbox}    
\usepackage{array}      
\usepackage{colortbl}
\usepackage{longtable}   
\usepackage{xcolor}
\usepackage{xpatch}
\usepackage{tikz}
\usepackage{cuted}
\usepackage{makecell}
\usepackage{hyperref}
\usepackage{booktabs}
\def\BibTeX{{\rm B\kern-.05em{\sc i\kern-.025em b}\kern-.08em
   T\kern-.1667em\lower.7ex\hbox{E}\kern-.125emX}}
\begin{document}

\title{OpenISAC: An Open-Source Real-Time Experimentation Platform for OFDM-ISAC}

\author{Zhiwen~Zhou, Chaoyue~Zhang, Xiaoli~Xu,~\IEEEmembership{Member,~IEEE}, and
        Yong~Zeng,~\IEEEmembership{Fellow,~IEEE}
        \vspace{-2em}
\thanks{This work was supported by the National Natural Science Foundation of China under Grant 62571116, by the Natural Science Foundation for Distinguished Young Scholars of Jiangsu Province under Grant BK20240070, and by the 2025 Yangtze River Delta Science and Technology Innovation Community Joint Research Project (Fundamental Research) under Grant 2025CSJZN01800. 

The authors are with the National Mobile Communications Research Laboratory, Southeast University, Nanjing 210096, China. Yong Zeng is also with the Purple Mountain Laboratories, Nanjing 211111, China (e-mail: \{zhiwen\_zhou, chaoyue\_zhang, xiaolixu, yong\_zeng\}@seu.edu.cn). (Corresponding author: Yong Zeng.)

The source code and a detailed user guide for OpenISAC are publicly available at \href{https://github.com/zhouzhiwen2000/OpenISAC}{https://github.com/zhouzhiwen2000/OpenISAC}.

Copyright (c) 20xx IEEE. Personal use of this material is permitted. However, permission to use this material for any other purposes must be obtained from the IEEE by sending a request to pubs-permissions@ieee.org.
}
}

\maketitle
\bstctlcite{IEEEexample:BSTcontrol}

\begin{abstract}
Integrated sensing and communication (ISAC) is envisioned to be one of the key usage scenarios for the sixth generation (6G) mobile communication networks. While significant progresses have been achieved for the theoretical studies, the further advancement of ISAC is hampered by the lack of accessible, open-source, and real-time experimental platforms. To address this gap, we introduce OpenISAC, a versatile and high-performance open-source platform for real-time ISAC experimentation.
OpenISAC utilizes orthogonal frequency division multiplexing (OFDM) waveform and implements crucial sensing functionalities, including both monostatic and bistatic delay-Doppler sensing. It is implemented as a fully open-source host-based real-time platform that combines high-speed C++ baseband processing with a Python interface for rapid prototyping, without relying on proprietary software or specialized FPGA development. The platform is built entirely on open-source software, leveraging the universal software radio peripheral (USRP) hardware driver (UHD) library, thus eliminating the need for any commercial licenses. It supports a wide range of software-defined radios, from the cost-effective USRP B200 series to the high-performance X400 series. The physical layer modulator and demodulator are implemented with C++ for high-speed processing, while the sensing data is streamed to a Python environment, providing a user-friendly interface for rapid prototyping and validation of sensing signal processing algorithms. It also includes practical features such as continuous-wave operation, flexible OFDM parameterization, and an optional over-the-air (OTA) synchronization capability for bistatic sensing without wired clock distribution. With flexible parameter selection and real-time communication and sensing operation, OpenISAC serves as a powerful and accessible tool for the academic and research communities to explore and innovate within the field of OFDM-ISAC.
\end{abstract}


\IEEEpeerreviewmaketitle
\section{Introduction}
Integrated sensing and communication (ISAC) has been identified as one of the key usage scenarios for future sixth-generation (6G) mobile networks \cite{sector2023framework}, poised to enable a wide range of applications like autonomous driving, robotics, environmental monitoring, smart cities, and emerging low-altitude economy services (e.g., unmanned aerial vehicle (UAV)-based logistics, inspection, and surveillance). By merging sensing and communication functionalities into a unified hardware platform and signal processing framework, ISAC can enhance spectral and energy efficiency while providing communication networks with native environmental awareness.

Orthogonal Frequency Division Multiplexing (OFDM) is a particularly attractive waveform for ISAC systems, given its widespread adoption in modern communications and its inherent suitability for performing delay-Doppler sensing \cite{dai2025tutorial,liu2024ofdm}. Crucially, OFDM is expected to remain the dominant waveform for 6G, as 3GPP has recently endorsed CP-OFDM and DFT-s-OFDM as the primary waveforms for the 6G radio interface \cite{3gpp_ran1_122_report}. Despite significant theoretical advancements, the practical exploration and validation of novel OFDM ISAC technologies are currently hampered by a critical bottleneck: the lack of accessible, open-source, and real-time experimental platforms. The development of such platforms is crucial for bridging the gap between theory and practice, enabling researchers to validate algorithms with real-world data and accelerate the innovation cycle.

A comparison of existing ISAC experimental platforms is presented in Table \ref{tab:platform_comparison}. Early research efforts have led to several prototype systems demonstrating the feasibility of OFDM-based ISAC. For instance, an early testbed based on Universal Software Radio Peripherals (USRPs) was proposed in \cite{braun2012usrp} to demonstrate OFDM radar functionalities. This system adopts a capture-and-process workflow where data is recorded and subsequently processed in MATLAB, resulting in a relatively low update rate. Similarly, the millimeter-wave mobile sensing platform in \cite{barneto2022millimeter} relies on RF measurements and post-processing to validate environment mapping algorithms. The full-duplex OFDM radar platform in \cite{barneto2019fullduplex} uses 5G NR waveforms and mainly focuses on analog- and digital-domain self-interference (SI) cancellation. Its reported 40~MHz NR experiment is based on playback and recording of a 20~ms waveform using vector signal transceivers (VSTs), rather than real-time CW operation. The potential of opportunistic use of signals was explored in \cite{5494565}, which presented a passive bistatic radar system using commercial WiFi signals. While demonstrating bistatic operation, this system relies on a wired connection to the access point for the reference signal, constraining its applications. To achieve real-time performance, several high-end platforms have been developed using Field Programmable Gate Arrays (FPGAs). Notable examples include the JCR70 millimeter-wave platform \cite{kumari2021jcr70} and the networking-based ISAC testbed \cite{ji2023networking}, which support wide bandwidths and real-time processing. However, similar to our previous work \cite{10233570}, these implementations typically rely on National Instruments LabVIEW and high-end FPGA devices, which entail significant hardware and software costs that limit their accessibility to the wider academic and small enterprise community. Furthermore, recent commercial 5G-A field trials \cite{zte2024lowaltitude, unicom2023vehicle, cmcc2023whitepaper} have demonstrated impressive ISAC capabilities in large-scale networks, but these proprietary solutions remain inaccessible to independent researchers.

\begin{table*}[!t]
\caption{Comparison of Existing ISAC Experimental Platforms}
\label{tab:platform_comparison}
\centering
\footnotesize
\renewcommand{\arraystretch}{1.35}
\resizebox{\textwidth}{!}{%
\begin{tabular}{|c|c|c|c|c|c|c|c|c|}
\hline
\textbf{Platform / Work} & \textbf{Implementation} & \textbf{Cost} & \textbf{Real-Time} & \textbf{\makecell{Transmission\\Scheme}} & \textbf{Bandwidth} & \textbf{\makecell{Bistatic\\Sensing}} & \textbf{\makecell{Bistatic\\Sync}} & \textbf{\makecell{Open\\Source}} \\ \hline
OFDM Radar Testbed \cite{braun2012usrp} & \makecell{Host\\(MATLAB)} & Medium & Yes & Packet & \makecell{Flexible\\(5-20~MHz)} & No & N/A & No \\ \hline
mmWave Mobile Sensing \cite{barneto2022millimeter} & \makecell{RF Measurements \&\\Post Processing} & High & No & \makecell{CW\\(5G NR)} & 400~MHz & No & N/A & No \\ \hline
Passive WiFi Radar \cite{5494565} & \makecell{Host\\(MATLAB)} & Medium & No & \makecell{Packet\\(WiFi)} & 20~MHz & Yes & Wired & No \\ \hline
JCR70 \cite{kumari2021jcr70} & FPGA/LabVIEW & High & Yes & Packet & 2~GHz & No & N/A & No \\ \hline
Networking Based ISAC \cite{ji2023networking} & FPGA/LabVIEW & High & Yes & \makecell{CW\\(5G NR)} & \makecell{800~MHz\\(100~MHz\\for sensing)} & No & N/A & No \\ \hline
mmWave OFDM ISAC \cite{10233570} & FPGA/LabVIEW & High & Yes & Packet & 100~MHz & No & N/A & No \\ \hline
Full-Duplex OFDM Radar \cite{barneto2019fullduplex} & \makecell{VST\\Playback/Record} & High & No & \makecell{Playback/Record\\(5G NR)} & 40~MHz & No & N/A & No \\ \hline
Commercial 5G-A \cite{zte2024lowaltitude, unicom2023vehicle, cmcc2023whitepaper} & \makecell{Commercial\\BS} & High & Yes & \makecell{CW\\(5G NR)} & 100-400~MHz & \makecell{Some\\\cite{cmcc2023whitepaper}} & Networked & No \\ \hline
OAI-based 5G ISAC \cite{carbonara2025oai}, \cite{carbonara2026downlink} & \makecell{Host\\(OAI \& SDR)} & Medium & Yes & \makecell{CW\\(5G NR)} & \makecell{Flexible\\(Sub-6 GHz)} & No & N/A & No \\ \hline
OpenWiFi \cite{9128614} & FPGA (ZYNQ) & Low & Yes & \makecell{Packet\\(WiFi)} & 20~MHz & No & N/A & Yes \\ \hline
ESPARGOS \cite{10739065} & MCU (ESP32) & Low & Yes & \makecell{Packet\\(WiFi)} & 40~MHz & \makecell{AoA\\Only} & N/A & Partial \\ \hline
\textbf{OpenISAC} & \textbf{\makecell{Host\\(C++ \& Python)}} & \textbf{Flexible} & \textbf{Yes} & \textbf{CW} & \textbf{\makecell{Flexible\\(Depends on\\PC \& USRP)}} & \textbf{Yes} & \textbf{OTA} & \textbf{Yes} \\ \hline
\end{tabular}%
}
\vspace{-1em}
\end{table*}

Recent academic efforts have also started to explore standards-compliant SDR-based ISAC. In particular, \cite{carbonara2025oai} demonstrates 5G-NR-compliant sensing experiments using OpenAirInterface (OAI) and SDR hardware. The OAI-based downlink ISAC testbed in \cite{carbonara2026downlink} further investigates and compares the role of control signals for multi-target detection while experimentally validating the testbed using a channel emulator. Such platforms are valuable for standard-aligned prototyping, but they inherit the frame-structure and waveform constraints of the underlying cellular stack, which limits the flexibility of continuous sensing-oriented waveform redesign.
NVIDIA Sionna and the Sionna Research Kit are also important open-source tools for wireless communication research, including ray-tracing-based channel modeling and SDR-enabled communication prototyping~\cite{sionna}. However, they primarily target communication-system simulation and prototyping rather than a complete end-to-end ISAC sensing pipeline. In contrast, OpenISAC provides a compact and generic OFDM-ISAC implementation without a complex protocol stack, allowing researchers to configure waveform parameters and the time-frequency resource grid more flexibly for ISAC experiments.

At the signal-processing level, OTA synchronization and asynchronous bistatic ISAC have already been studied in the literature \cite{brunner2025bistatic,wu2024clock,ding2025bistatic}. OpenISAC includes this capability as a practical feature for wire-free bistatic operation.

In recent years, several open-source ISAC platforms have emerged, primarily leveraging the WiFi ecosystem. OpenWiFi is a notable community-driven project that has incorporated sensing functionalities into its open-source IEEE 802.11-compliant design\cite{9128614}. However, the OpenWiFi architecture relies heavily on FPGAs for baseband processing, which makes accessing and manipulating low-level signal data difficult for researchers focusing on physical-layer innovations. Another representative platform, ESPARGOS, provides a low-cost ESP32-based antenna array for spatial-domain and channel state information (CSI)-based sensing\cite{10739065}. While valuable, ESPARGOS is not fully open-source (its hardware and firmware are closed) and primarily targets spatial-domain and CSI-based sensing rather than the delay-Doppler processing that is central to many ISAC applications. More broadly, platforms such as OpenWiFi and ESPARGOS are designed for WiFi compliance, adhering to packet-based transmission schemes with relatively fixed frame structures and parameters, which complicates sensing operations that require continuous observation (e.g., Doppler and micro-Doppler sensing across multiple frames). In addition, most existing systems are predominantly monostatic, assuming co-located transmit and receive nodes and providing limited support for bistatic or multistatic operation.

To address the above challenges, in this paper we propose and develop OpenISAC, a versatile, high-performance, and fully open-source platform for real-time OFDM ISAC experimentation. OpenISAC adopts a continuous-wave (CW) OFDM transmission scheme with a flexible frame structure, which greatly simplifies cross-frame sensing operations such as Doppler and micro-Doppler extraction. The entire physical layer is implemented on a host CPU using the USRP Hardware Driver (UHD) C++ API, and the resulting data streams are exposed to a user-friendly Python environment for advanced signal processing. This hybrid C++/Python architecture combines real-time performance with a highly accessible interface for rapid prototyping of sensing and communication algorithms. The platform supports a wide range of software-defined radios, from cost-effective USRP B200-series devices to high-performance X400-series hardware, and performs communication and sensing in real time. OpenISAC provides a rather complete communication function, allowing sensing algorithms to be evaluated under realistic communication constraints such as data symbols, pilots, synchronization fields, frame structure, and receiver timing behavior. It can also serve as a compact real-time OFDM testbed for communication algorithms that depend on real channel statistics, while enabling evaluation of processing latency and real-time feasibility. In addition, OpenISAC includes practical support for bistatic operation, including an OTA synchronization option when a wired synchronization link is unavailable. By providing an accessible, flexible, and powerful tool, the proposed OpenISAC aims to empower the academic and research communities to explore and innovate within the burgeoning field of ISAC.

The main contributions of this paper are as follows:
\begin{itemize}

\item \textbf{Open-Source Real-Time Implementation:} We present the design and public release of OpenISAC, a fully open-source, real-time ISAC platform. Its host-based architecture, combining C++ for high-speed processing with Python for user-facing development, provides both high performance and broad accessibility, thereby eliminating dependencies on proprietary software or specialized FPGA expertise. The source code and a detailed user guide are publicly available at \href{https://github.com/zhouzhiwen2000/OpenISAC}{https://github.com/zhouzhiwen2000/OpenISAC}.

\item \textbf{Practical Bistatic Operation:} OpenISAC includes support for bistatic sensing in practical deployments, including an OTA synchronization pipeline that can be used when a wired synchronization link is unavailable. This feature is integrated into the real-time software stack together with carrier-frequency tracking, sampling-clock tracking, fractional delay refinement, and sensing-side timing compensation.

\item \textbf{Flexible Physical Layer Design:} We demonstrate a highly flexible physical layer that supports both continuous-wave transmission and fully customizable OFDM parameters. This empowers researchers to design and test novel waveforms optimized for sensing, breaking free from the rigid constraints of standard-compliant, packet-based systems.

\item \textbf{Experimental Validation:} We experimentally validate real-time monostatic and bistatic sensing, communication performance, and host-side runtime behavior, demonstrating that OpenISAC is a practical and accessible platform for OFDM-ISAC experimentation.

\end{itemize}

The remainder of this paper is organized as follows. Section~\ref{sec:signal_model} introduces the signal model. Section~\ref{sec:sensing} details the signal processing framework for communication as well as monostatic and bistatic sensing, including the practical OTA synchronization procedure used for bistatic operation. Section~\ref{sec:sysarch} describes the OpenISAC system architecture, covering its hardware and software implementation. Section~\ref{sec:exp_results} presents experimental results validating the platform's performance, and Section~\ref{conclusion} concludes the paper.

\textit{Notation}: Italic, bold lower-case, and bold upper-case symbols denote scalars, vectors, and matrices, respectively. $(\cdot)^T$, $(\cdot)^H$, and $(\cdot)^*$ denote transpose, Hermitian transpose, and complex conjugate; $\mathcal{CN}(0,\sigma^2)$ denotes a circularly symmetric complex Gaussian (CSCG) random variable; and $\mathrm{Re}\{\cdot\}$, $\mathrm{Im}\{\cdot\}$, and $\mathrm{sgn}(\cdot)$ denote the real part, imaginary part, and sign function, respectively.

\section{System Model}\label{sec:signal_model}
\begin{figure}[!htbp]
\centering
\includegraphics[width=0.35\textwidth]{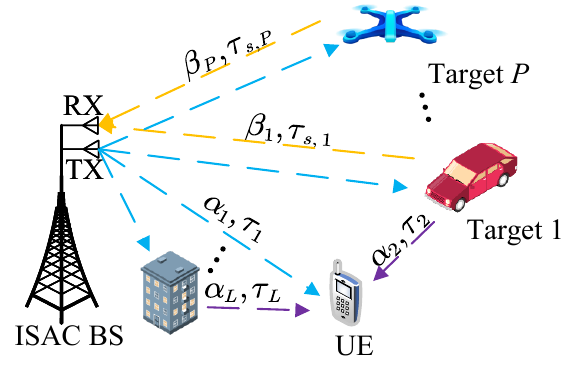}
\caption{An illustration of the system model for the developed OpenISAC platform, one communication UE and $P$ sensing targets.}
\label{fig:model}
\end{figure}
We consider an OFDM-based ISAC system comprising one base station (BS) and one communication user equipment (UE). The prototype presented in this paper follows a SISO configuration, with one BS transmit antenna, one BS monostatic receive antenna, and one UE receive antenna. The communication link is downlink-only, with no uplink transmission in the current implementation. The system has three simultaneous objectives: 1) downlink communication with the UE; 2) monostatic sensing utilizing the backscattered echoes received at the BS; and 3) bistatic sensing utilizing the signals received at the UE. BS-side monostatic sensing does not require the presence of the UE; the UE is needed only for downlink communication reception and UE-side bistatic sensing. In this paper, we refer to the resolvable objects in the BS-BS monostatic channel as \textit{sensing targets}, and the resolvable objects in the BS-UE channel as \textit{scatterers}. It is worth noting that these two sets can have overlapping elements, meaning a physical object may act as both a sensing target and a scatterer.

As shown in Fig.~\ref{fig:model}, the BS-UE channel can be written as
\begin{equation}
\abovedisplayshortskip=2pt
\belowdisplayshortskip=2pt
\abovedisplayskip=2pt
\belowdisplayskip=2pt
\begin{aligned}
{h}_{\mathrm{UE}}\left( t,\tau \right) =\sum_{l=1}^L{\alpha _l\delta \left( \tau -\tau _l-\tau _d \right) e^{j2\pi \left( f_{D,l}+\Delta f_c \right) t}},
\end{aligned}
\label{Channel_BS_UE}
\end{equation}
where $L$ is the number of multi-path components (MPCs), with $\alpha_l$, $\tau_l$ and $f_{D,l}$ denoting the complex-valued scattering coefficient, delay and Doppler shift for path $l$, respectively. The term $\tau_d$ denotes the timing offset between the BS and the UE and $\Delta f_c$ is the carrier-frequency offset. Without loss of generality, we assume that the strongest path is $l=1$. Given that the bistatic sensing receiver operates over the same time-frequency (TF) resources and observes the same channel as the communication receiver, we employ the channel model in \eqref{Channel_BS_UE} for both BS-UE communication and bistatic sensing. Specifically, for bistatic sensing, the MPCs with low or zero Doppler shifts are treated as clutter, while dynamic paths are identified as valid dynamic scatterers.

The monostatic sensing channel can be written as
\begin{equation}
\abovedisplayshortskip=2pt
\belowdisplayshortskip=2pt
\abovedisplayskip=2pt
\belowdisplayskip=2pt
\begin{aligned}
{h}_{\mathrm{BS}}\left( t,\tau \right) &= \sum_{p=1}^{P+C}{\beta _{p}\delta \left( \tau -\tau _{s,p} \right) e^{j2\pi f_{D,s,p} t}},
\end{aligned}
\label{Channel_BS_BS}
\end{equation}
where $P$ is the number of targets and $C$ is the number of clutter echoes. The clutter components $p=P+1, \dots, P+C$ are assumed to have low or zero Doppler shifts. Unlike the bistatic BS-UE link, the monostatic sensing channel model in \eqref{Channel_BS_BS} does not include the timing offset $\tau_d$ or the carrier-frequency offset $\Delta f_c$. This is because the monostatic transmitter and receiver are co-located at the BS and share the same reference clock. For a specific reflection $p$ with radar cross section (RCS) $\sigma_{\mathrm{RCS},p}$, range $d_{s,p}$, and radial velocity $v_{s,p}$, the complex scattering coefficient, round-trip delay, and two-way Doppler shift are modeled as
\begin{equation}
\abovedisplayshortskip=2pt
\belowdisplayshortskip=2pt
\abovedisplayskip=2pt
\belowdisplayskip=2pt
\begin{aligned}
\beta_p=\sqrt{\frac{c^{2}\sigma_{\mathrm{RCS},p}}{(4\pi)^3 d_p^{4} f_c^{2}}} e^{j\phi_p},\quad
\tau_{s,p}=\frac{2 d_p}{c},\quad
f_{D,s,p}=\frac{2 v_p f_c}{c},
\end{aligned}
\label{Params_Targets}
\end{equation}
where $c$ is the speed of light, $f_c$ is the carrier frequency, and $\phi_p$ denotes the phase shift induced by the reflection.

\subsection{Transmission Scheme}
\begin{figure}[!htbp]
\centering
\includegraphics[width=0.4\textwidth]{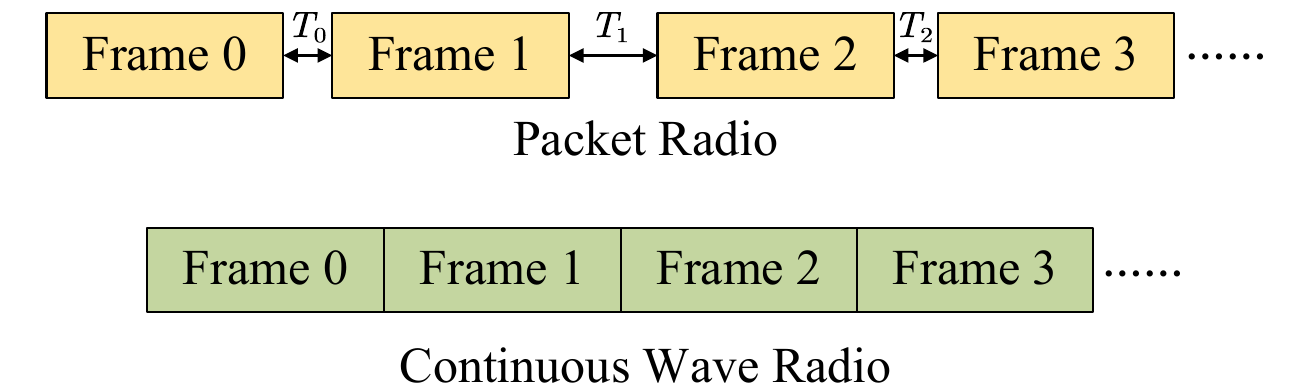}
\caption{Comparison between packet radio and CW radio.}
\label{fig:packet_vers_cw}
\end{figure}

As shown in Fig.~\ref{fig:packet_vers_cw}, mainstream digital radios fall into two transmission styles: packet radio (e.g., Wi-Fi) and CW waveforms (e.g., long term evolution (LTE) or new radio (NR)). In packet radio, carrier-sense multiple access control (MAC) and traffic-driven burstiness result in irregular idle periods between frames, illustrated as unequal intervals $T_0, T_1, \dots$ in the figure. This randomness makes the frame timing difficult to control. In Wi-Fi-based passive/bistatic sensing, this translates into a variable pulse-repetition interval (PRI) that distorts the Doppler spectrum and raises unpredictable sidelobes\cite{5960576}. In CSI-based Wi-Fi sensing, using ordinary data traffic yields uneven sampling over time and often too few packets to sustain coherent accumulation, directly hurting Doppler resolution and micro-Doppler fidelity\cite{9814523,he2023sencom}.
In addition, commodity Wi-Fi sensing platforms typically expose per-packet timestamps at millisecond-level resolution\cite{openbsd-ieee80211-radiotap,esp32-forum-35385}, which is far coarser than the nanosecond-level interval information desirable for high-accuracy inter-frame phase processing.
Overall, without inter-frame processing, the curtailed coherent accumulation lowers Doppler resolution and micro-Doppler fidelity; with inter-frame processing, the unequal spacing and inaccurate inter-frame interval readings often induce Doppler-domain artifacts and estimation errors.

By contrast, for CW radios such as LTE or NR, sensing OFDM symbols can be chosen with deterministic and uniform intervals, enabling longer, cleaner coherent integration and more flexible Doppler/micro-Doppler processing.
Guided by these considerations, OpenISAC adopts the CW scheme to enable more accurate and flexible Doppler sensing.
In OpenISAC, the BS transmits continuous OFDM frames, which can be expressed as
\begin{equation}
\abovedisplayshortskip=2pt
\belowdisplayshortskip=2pt
\abovedisplayskip=2pt
\belowdisplayskip=2pt
\begin{aligned}
s\left( t \right) =\sum_{\gamma =0}^{\infty}{s_{\gamma}\left( t-\gamma T_F \right)},
\end{aligned}
\label{Signal}
\end{equation}
where $T_F=MT_O$ is the frame duration, with $M$ and $T_O$ denoting the number of OFDM symbols per frame and the OFDM symbol duration, respectively. The transmit signal of the $\gamma$th frame  can be written as
\begin{equation}
\abovedisplayshortskip=2pt
\belowdisplayshortskip=2pt
\abovedisplayskip=2pt
\belowdisplayskip=2pt
\begin{aligned}
s_{\gamma}\left( t \right) =\sum_{m=0}^{M-1}{\!\sum_{n=0}^{N-1}{b_{n,m,\gamma}e^{j2\pi n\Delta f\left( t\!-\!mT_O\!-\!T_\mathrm{CP} \right)}\mathrm{rect}\!\left( \!\frac{t\!-\!mT_O}{T_O}\! \right) \!}},
\end{aligned}
\label{Signal_frame}
\end{equation}
where $\Delta f$ is the subcarrier spacing, $T_\mathrm{CP}$ is the duration of the cyclic prefix (CP). The duration of the OFDM symbol with CP can then be expressed as $T_O=T+\!T_\mathrm{CP}$, with $T\triangleq1/\Delta f$. The matrix $\boldsymbol{B}_{\gamma}\triangleq [b_{n,m,\gamma}]_{n=0,m=0}^{N-1,M-1}\in\mathbb{C}^{N\times M}$ consists of the $NM$ constellation symbols for the $\gamma$th frame. The element ${b_{n,m,\gamma}}$ denotes the symbol modulated onto the $n$th subcarrier and $m$th OFDM symbol, normalized such that $\mathbb{E} \left\{ \left|b_{n,m,\gamma} \right|^2 \right\} =P_\mathrm{Tx}/N$, where $P_\mathrm{Tx}$ represents the average transmit power.

\subsection{Frame Structure and Sensing Signal Model}
\begin{figure}[htbp]
    \centering
    \includegraphics[width=0.8\linewidth]{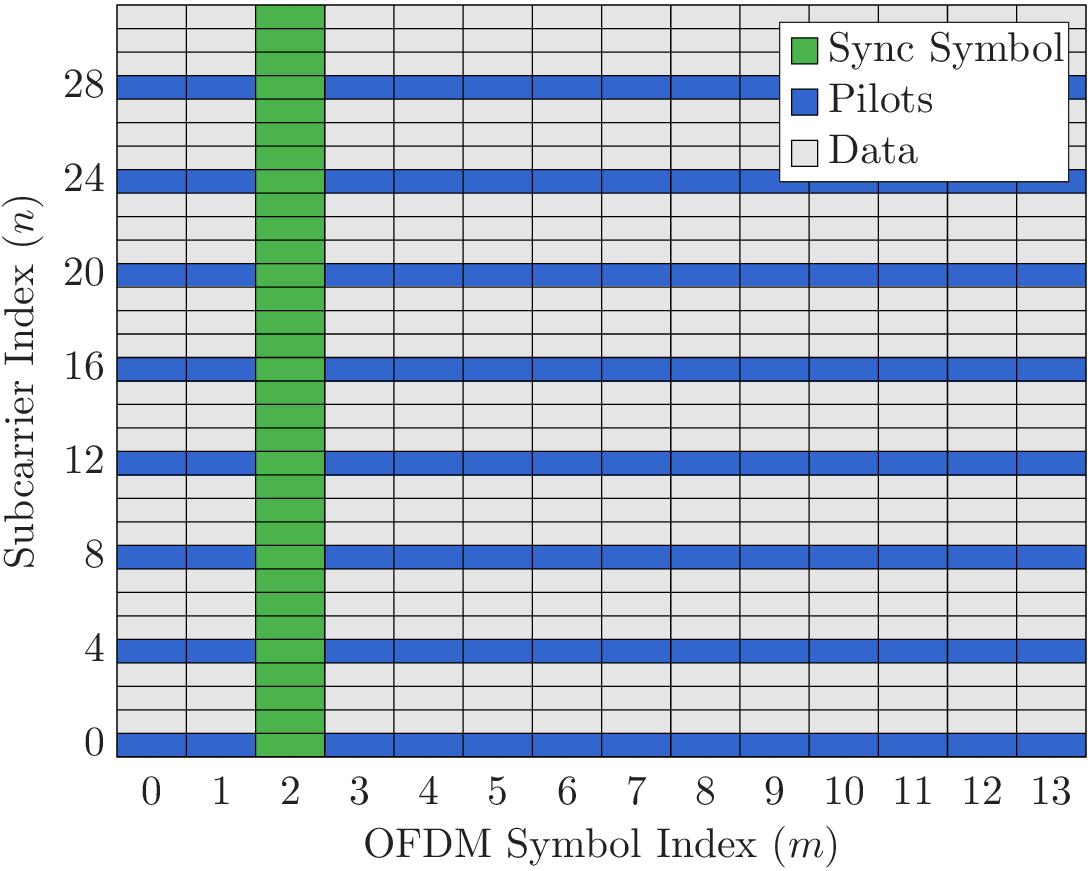}
    \caption{Example of the OFDM frame structure with $M=14$, $N=32$, $m_{\mathrm{sync}}=2$, and $\mathcal{P}=\{0, 4, \dots, 28\}$.}
    \label{fig:frame_structure}
\end{figure}

In order for the UE to synchronize with the BS and to obtain reliable per-symbol channel estimates, as illustrated in Fig.~\ref{fig:frame_structure}, OpenISAC reserves one mandatory OFDM symbol in each frame, indexed by $m_{\mathrm{sync}}\in\{0,\ldots,M-1\}$, as a full-band ZC synchronization symbol and embeds pilots on a fixed set of subcarriers for all data-bearing OFDM symbols. Since OpenISAC is intended as a lightweight OFDM-ISAC PHY for experimentation rather than a Wi-Fi-, LTE-, or NR-compliant implementation, these fields are designed as configurable reference signals instead of standard PRACH or other standardized access signals. For deployments with larger initial CFO, the frame can also reserve two optional acquisition fields: a duplicate full-band ZC synchronization symbol at $m_{\mathrm{sync}}-1$ and a repeated CFO training field at $m_{\mathrm{sync}}+1$. Specifically, let $\mathcal{P}\subset\{0,\ldots,N-1\}$ denote the set of pilot subcarrier indices, and let $\mathcal{D}\triangleq\{0,\ldots,N-1\}\setminus\mathcal{P}$ denote the data-subcarrier set.
Let $\mathcal{S}_{\mathrm{sec}}=\{m_{\mathrm{sync}}-1\}$ when the optional second synchronization symbol is enabled and $\mathcal{S}_{\mathrm{sec}}=\emptyset$ otherwise, with $m_{\mathrm{sync}}\geq1$ required when it is enabled. Define $\mathcal{S}_{\mathrm{ZC}}\triangleq\{m_{\mathrm{sync}}\}\cup\mathcal{S}_{\mathrm{sec}}$. Similarly, let $\mathcal{S}_{\mathrm{CFO}}=\{m_{\mathrm{sync}}+1\}$ when the optional CFO training field is enabled and $\mathcal{S}_{\mathrm{CFO}}=\emptyset$ otherwise.
Let $d_{n,m,\gamma}$ denote the quadrature phase shift keying (QPSK) data symbols mapped from low-density-parity-check (LDPC)-encoded and scrambled bits with alphabet $\mathcal{A}_{\text{QPSK}}=\big\{\tfrac{1}{\sqrt{2}}(\pm 1 \pm j)\big\}$. LDPC encoding/decoding is provided by the AFF3CT library \cite{cassagne2019aff3ct}. Let $c_n^{\mathrm{CFO}}$ denote the frequency-domain symbols of the repeated CFO training field. With these definitions, the constellation symbols in \eqref{Signal_frame} are given by
\begin{equation}
b_{n,m,\gamma}=
\begin{cases}
z_n, & \mathrm{if}\, m\in\mathcal{S}_{\mathrm{ZC}}\ \mathrm{and}\ n\in\{0,\ldots,N-1\},\\[2pt]
c_n^{\mathrm{CFO}}, & \mathrm{if}\, m\in\mathcal{S}_{\mathrm{CFO}}\ \mathrm{and}\ n\in\{0,\ldots,N-1\},\\[2pt]
z_n, & \mathrm{if}\, m\notin\mathcal{S}_{\mathrm{ZC}}\cup\mathcal{S}_{\mathrm{CFO}}\ \mathrm{and}\ n \in \mathcal{P},\\[2pt]
d_{n,m,\gamma}, & \mathrm{if}\, m\notin\mathcal{S}_{\mathrm{ZC}}\cup\mathcal{S}_{\mathrm{CFO}}\ \mathrm{and}\ n \in \mathcal{D},
\end{cases}
\label{eq:bnm_mapping}
\end{equation}
where $\{z_n\}_{n=0}^{N-1}$ is a length-$N$ Zadoff-Chu (ZC) sequence with root $q$ coprime with $N$. The ZC synchronization symbols are full-band OFDM symbols used for time/frequency synchronization and coarse channel acquisition. In the compact one-ZC-symbol configuration, $\mathcal{S}_{\mathrm{sec}}=\mathcal{S}_{\mathrm{CFO}}=\emptyset$, which reduces synchronization/pilot overhead and keeps all OFDM symbols, including the full-band ZC synchronization symbol, available as sensing-symbol candidates. When the optional CFO training field is enabled, that field is reserved for acquisition and is excluded from sensing-symbol selection. As shown in Fig.~\ref{fig:frame_structure}, for all data-bearing OFDM symbols only the pilot subcarriers $n\in\mathcal{P}$ carry ZC entries and the remaining subcarriers $n\in\mathcal{D}$ carry QPSK data. When no downlink payload is available, these data resources are filled with random QPSK padding symbols, and optional dedicated sensing time-frequency resources can also be configured. After this frequency-domain mapping, the transmitter forms the TF resource grid $\boldsymbol{B}_{\gamma}$, stores a copy for sensing-side processing (e.g., range-Doppler processing), and then applies the inverse fast Fourier transform (IFFT) followed by CP insertion to generate $s_\gamma(t)$ in \eqref{Signal_frame}.

The signal received by the monostatic radar receiver can be written as
\begin{equation}
\begin{aligned}
y_{\mathrm{BS}}\left( t \right) &=\int_{-\infty}^{\infty}{h_{\mathrm{BS}}\left( t,\tau \right) s\left( t-\tau \right) d\tau}+z_{\mathrm{BS}}\left( t \right) 
\\
&=\sum_{p=1}^{P+C}{\beta _ps\left( t -\tau _{s,p} \right) e^{j2\pi f_{D,s,p}t}}+z_{\mathrm{BS}}\left( t \right),
\end{aligned}
\label{eq:mono_rx}
\end{equation}
where $z_{\mathrm{BS}}\left( t \right)\sim \mathcal{CN}(0, \sigma^2)$ represents the additive white Gaussian noise, with $\sigma^2 = B N_0  $ denoting the noise power. Here, $B = N \Delta f$ is the signal bandwidth, and $N_0$ is the double-sided noise power spectral density. The receiver then samples the received signal with exactly the same sampling frequency as the transmitter, i.e., $f_s=B=N\Delta f$. The received signal after sampling can be expressed as
\begin{equation}
\begin{aligned}
y_{\mathrm{BS}}\left[ k \right] &=y_{\mathrm{BS}}\left( kT_s \right) 
\\
&=\sum_{p=1}^{P+C}{\beta _ps\left( kT_s-\tau _{s,p} \right) e^{j2\pi f_{D,s,p}kT_s}}+z_{\mathrm{BS}}\left[ k \right],
\end{aligned}
\label{eq:mono_rx_samp}
\end{equation}
where $T_s=1/B$ is the sampling interval. 
The signal received by the UE can be written as
\begin{equation}
\begin{aligned}
y_{\mathrm{UE}}\left( t \right) =\int_{-\infty}^{\infty}{h_{\mathrm{UE}}\left( t,\tau \right) s\left( t-\tau -\tau _d \right) d\tau}+z_{\mathrm{UE}}\left( t \right) 
\\
=\sum_{l=1}^L{\alpha _ls\left( t -\tau _l - \tau_d \right) e^{j2\pi \left( f_{D,l}+\Delta f_c \right) t}}+z_{\mathrm{UE}}\left( t \right) ,
\end{aligned}
\label{eq:ue_rx}
\end{equation}
where $z_{\mathrm{UE}}\left( t \right)\sim \mathcal{CN}(0, \sigma^2)$ is the Gaussian noise. Due to sampling frequency offset (SFO), the UE samples the signal with a slightly different sampling interval $T_{s,\mathrm{UE}}=T_s-\Delta T_s$:
\begin{equation}
\begin{aligned}
y_{\mathrm{UE}}\left[ k \right] \!=\!y_{\mathrm{UE}}\left( k\left( T_s-\Delta T_s \right) \right) \!=\!\sum_{l=1}^L{y_l\left[ k \right]}+z_{\mathrm{UE}}\left[ k \right] ,
\end{aligned}
\label{eq:bi_rx_samp}
\end{equation}
where $y_l\left[ k \right] $ is expressed as
\begin{equation}
\begin{aligned}
y_l\left[ k \right] \!=\!\alpha _ls\left( kT_s\!-\!\tau _l\!-\!\tau _d\!-\!k\Delta T_s \right) e^{j2\pi \left( f_{D,l}+\Delta f_c \right) k\left( T_s-\Delta T_s \right)}.
\end{aligned}
\label{eq:bi_rx_samp_path}
\end{equation}
Since the Doppler shift, carrier-frequency offset (CFO) and the sampling interval offset (SIO) are relatively small, the cross term $(f_{D,l}+\Delta f_c)\Delta T_s$ can be ignored. Thus, \eqref{eq:bi_rx_samp_path} can be approximated as
\begin{equation}
\begin{aligned}
y_l\left[ k \right] \!\approx \!\alpha _ls\left( kT_s\!-\!\tau _l\!-\!\tau _d\!-\!k\Delta T_s \right) e^{j2\pi \left( f_{D,l}+\Delta f_c \right) kT_s}.
\end{aligned}
\label{eq:bi_rx_samp_path_approx}
\end{equation}

With the received-signal models in \eqref{eq:mono_rx_samp} and \eqref{eq:bi_rx_samp} established, we next describe the baseband signal processing for OpenISAC.
Specifically, the BS exploits its received samples $y_{\mathrm{BS}}[k]$ for monostatic sensing, whereas the UE processes $y_{\mathrm{UE}}[k]$ for both downlink communication reception and BS-UE bistatic sensing.

\section{Signal Processing for OpenISAC}  \label{sec:sensing}
This section introduces the signal processing procedures for the three core functionalities of the proposed platform: monostatic sensing at the BS, BS-UE communication, and BS-UE bistatic sensing.
\subsection{Monostatic Sensing} \label{subsec:monosensing}
\begin{figure}[!htbp]
\centering
\includegraphics[width=0.45\textwidth]{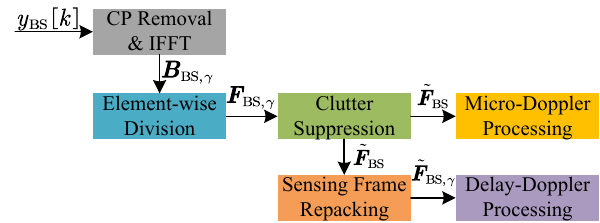}
\caption{Signal processing procedure for monostatic sensing.}
\label{fig:monosensing}
\end{figure}
As shown in Fig.~\ref{fig:monosensing}, the monostatic sensing pipeline starts from the received samples $y_{\mathrm{BS}}[k]$, which are first mapped onto the OFDM TF resource grid by CP removal and FFT. The resulting symbols are then element-wise divided by the transmitted modulation symbols $\mathbf{B}_{\mathrm{BS},\gamma}$ to remove the influence of random communication symbols and obtain the TF channel matrices $\mathbf{F}_{\mathrm{BS},\gamma}$. Next, the resulting matrices are concatenated and fed into a clutter-suppression stage, where a slow-time high-pass filter is applied to remove static components, yielding $\tilde{\mathbf{F}}_{\mathrm{BS}}$. From here, two analysis branches are used: (a) a micro-Doppler branch that directly exploits $\tilde{\mathbf{F}}_{\mathrm{BS}}$ as a continuous slow-time stream to reveal fine motion signatures; and (b) a sensing-frame repacking step that reorganizes $\tilde{\mathbf{F}}_{\mathrm{BS}}$ into sensing-optimized frames $\tilde{\mathbf{F}}_{\mathrm{BS},\gamma}$, which are then processed by a delay-Doppler module to form range-Doppler spectra.

\subsubsection{TF Grid Mapping \& Element-wise Division}
Under the assumption that the maximum delay $\tau _{s,\max}={\max}\left\{ \tau_{s,p} \right\}$ does not exceed the CP duration $T_\mathrm{CP}$, and that the maximum Doppler shift $f _{D,s,\max}={\max}\left\{ f_{D,s,p} \right\}$ is smaller than $\Delta f/10$~\cite{braun2014ofdm}, then after CP removal, the received signal of the $\gamma$th frame can be rearranged into $N\times M$ receive matrix $ \boldsymbol{B}_{\mathrm{BS},\gamma}$ using the fast Fourier transform (FFT)~\cite{dai2025tutorial},
\begin{equation}
\begin{aligned}
\left( \!\boldsymbol{B}_{\mathrm{BS},\gamma}\! \right) _{n,m}\!&=\!b_{n,m,\gamma}\!\sum_{p=1}^{P+C}{\beta _pe^{j2\pi (f_{D,s,p}\left( m+\gamma M \right) T_O\!-\!n\Delta f\tau _{s,p})}\!}
\\
&+\!\left( \boldsymbol{Z}_{\mathrm{BS},\gamma} \right) _{n,m},
\end{aligned}
\label{eq:mono_rx_grid}
\end{equation}
where $\boldsymbol{Z}_{\mathrm{BS},\gamma}\in\mathbb{C}^{N\times M}$ is the noise matrix. For monostatic sensing, the transmitted resource grid $\boldsymbol{B}$ is known at the radar receiver. Thus, element-wise division can be employed to remove the influence of transmitted data symbols,
\begin{equation}
\begin{aligned}
&\left( \boldsymbol{F}_{\mathrm{BS},\gamma} \right) _{n,m}=\frac{\left( \boldsymbol{B}_{\mathrm{BS},\gamma} \right) _{n,m}}{b_{n,m,\gamma}}
\\
&=\sum_{p=1}^{P+C}{\beta_pe^{j2\pi (f_{D,s,p}\left( m+\gamma M \right) T_O\!-\!n\Delta f\tau _{s,p})}}+(\tilde{\boldsymbol{Z}}_{\mathrm{BS},\gamma})_{n,m},
\end{aligned}
\label{eq:mono_rx_grid_channel}
\end{equation}
where $(\tilde{ \boldsymbol{Z}}_{\mathrm{BS},\gamma} ) _{n,m}=\left( \boldsymbol{Z}_{\mathrm{BS},\gamma} \right) _{n,m}/b_{n,m,\gamma}$. In the following, the symbols after element-wise division are referred to as OFDM channel symbols.
In the taxonomy of correlation-based OFDM radar receivers in \cite{Mercier20}, this operation corresponds to a proximate reciprocal filter with circular correlation (PRF-CC). Since the ZC and QPSK symbols used in OpenISAC are constant-modulus, the PRF-CC operation in \eqref{eq:mono_rx_grid_channel} is equivalent to the corresponding proximate matched filter with circular correlation (PMF-CC), up to a common symbol-energy normalization.


\subsubsection{Clutter Suppression \& Sensing Frame Repacking}
Since continuous frames are transmitted, the OFDM channel symbols in different frames can be concatenated and accumulated as
\begin{equation}
\begin{aligned}
\left( \boldsymbol{F}_{\mathrm{BS}} \right) _{n,\gamma M+m}\triangleq \left( \boldsymbol{F}_{\mathrm{BS},\gamma} \right) _{n,m}.
\end{aligned}
\label{eq:mono_rx_grid_channel_accumulate}
\end{equation}
The resulting $\boldsymbol{F}_{\mathrm{BS}}$ can be expressed as
\begin{equation}
\begin{aligned}
\left( \boldsymbol{F}_{\mathrm{BS}} \right) _{n,m}\triangleq \sum_{p=1}^{P+C}{\beta_pe^{j2\pi (f_{D,s,p}mT_O\!-\!n\Delta f\tau _{s,p})}}+(\tilde{\boldsymbol{Z}}_{\mathrm{BS}})_{n,m}.
\end{aligned}
\label{eq:mono_rx_grid_channel_accumulate_2}
\end{equation}
Optionally, the OFDM channel symbols can be downsampled with factor $M_D$ to achieve tradeoff between sensing performance and computational complexity,
\begin{equation}
\begin{aligned}
( \grave{\boldsymbol{F}}_{\mathrm{BS}} ) _{n,m} = \left( \boldsymbol{F}_{\mathrm{BS}} \right) _{n,mM_D}.
\end{aligned}
\label{eq:mono_rx_grid_channel_downsamp}
\end{equation}
Then, we apply an improved moving target indication (MTI) procedure to suppress static (near-zero-Doppler) clutter. In essence, MTI applies a temporal high-pass filter along the slow-time index $m$, thereby creating a notch around zero Doppler. To obtain a narrow stopband at low computational cost, we adopt a causal IIR high-pass implementation,
\begin{equation}
\label{eq:mono_rx_grid_channel_IIR}
( \tilde{\boldsymbol{F}}_{\mathrm{BS}} )_{n,m}
= \frac{1}{a_0} \left( \sum_{i=0}^{I} b_i ( \grave{\boldsymbol{F}}_{\mathrm{BS}} )_{n,m-i}
- \sum_{j=1}^{J} a_j ( \tilde{\boldsymbol{F}}_{\mathrm{BS}} )_{n,m-j} \right),
\end{equation}
where $\{b_i\}_{i=0}^{I}$ and $\{a_j\}_{j=0}^{J}$ are the feedforward/feedback coefficients. For the clutter components indexed by $p=P+1, \dots, P+C$, which exhibit near-zero Doppler shifts (i.e., $f_{D,s,p} \approx 0$), the IIR high-pass response provides strong attenuation. In contrast, the sensing targets indexed by $p=1, \dots, P$ possess non-zero Doppler shifts and are thus largely preserved after filtering.

Compared with an FIR MTI filter of similar notch sharpness, the IIR design requires a significantly lower order, thereby reducing computational cost. Stability is ensured by placing all poles strictly inside the unit circle, and the notch cutoff should be set slightly above the maximum expected clutter Doppler. Our implementation allows customization of the filter coefficients. In the experiments, we used the MATLAB Filter Design Tool to design a Butterworth high-pass IIR filter with normalized stopband/passband edges $0.005/0.01$, stopband attenuation $80$~dB, passband ripple $1$~dB, and filter order $15$, and exported the coefficients in second-order-section form. Other high-pass IIR filters or classical MTI cancellers can also be used when a different clutter-notch width is desired. The released code provides a MATLAB script to convert the resulting second-order section (SOS) matrix and gain vector into C code.


After clutter rejection, the stream splits into two branches: one feeds micro-Doppler sensing, while the other performs delay-Doppler sensing. For delay-Doppler processing, sensing frame repacking is performed to construct the sensing frames,
\begin{equation}
\label{eq:mono_rx_grid_repack}
(\tilde{\boldsymbol{F}}_{\mathrm{BS},\gamma})_{n,0:M_s-1}=(\tilde{\boldsymbol{F}}_{\mathrm{BS}})_{n,\gamma M_s:\left( \gamma +1 \right) M_s-1},
\end{equation}
where $M_s$ is the number of OFDM channel symbols within a sensing frame. The sensing-frame length $M_s$ and the communication-frame length $M$ are independently configurable parameters, and $M_s$ can be freely designed to satisfy Doppler sensing requirements.
\subsubsection{Delay Doppler Processing}
Then, the periodogram can be calculated to obtain the delays and Doppler shifts of the targets,
\begin{equation}
\begin{aligned}
&\left(\mathrm{Per}_{\gamma}\right)_{k_{\tau},k_{f}}
=\frac{1}{N M_s}
\\
&\cdot\left|
\sum_{m=0}^{M_s-1}\sum_{n=0}^{N-1}
(\tilde{\boldsymbol{F}}_{\mathrm{BS},\gamma})_{n,m}\,
w[n,m]\,
e^{j2\pi \frac{n k_{\tau}}{N_{\mathrm{Per}}}}
e^{-j2\pi \frac{m k_{f}}{M_{\mathrm{Per}}}}
\right|^2,
\end{aligned}
\label{eq:Per}
\end{equation}
where $w[n,m]$ is a window function and $N_{\mathrm{Per}}\geq N$, $M_{\mathrm{Per}} \geq M_s$ are the delay and Doppler grid sizes. Equation~\eqref{eq:Per} is efficiently implemented by FFT/IFFT, and the peak locations provide the target delay and Doppler estimates
\begin{equation}
\label{eq:delay_Doppler}
\hat{\tau}=\frac{\hat{k}_{\tau}}{N_{\mathrm{Per}}\Delta f}, \quad\hat{f}_{D}=\frac{\hat{k}_{f}}{M_{\mathrm{Per}}M_D T_O},
\end{equation}
respectively.
\subsubsection{Micro-Doppler Processing}
After clutter suppression, micro-Doppler analysis operates directly on the slow-time stream of a selected delay bin. First, $N$-point IFFTs along the subcarrier index $n$ produce the delay-time matrix
\begin{equation}
\label{eq:md_delay_time}
\left(\boldsymbol{R}_{\mathrm{BS}}\right)_{k_{\tau},m}
=\frac{1}{N}\sum_{n=0}^{N-1} \left(\tilde{\boldsymbol{F}}_{\mathrm{BS}}\right)_{n,m}\,e^{\,j2\pi \frac{n k_{\tau}}{N}},
\end{equation}
where $\tilde{\boldsymbol{F}}_{\mathrm{BS}}$ is given by the MTI stage. Then a working delay bin $k_{\tau}^\star$ (e.g., strongest echo) is selected to form the slow-time sequence $r_{\mathrm{BS}}[m]\triangleq\left(\boldsymbol{R}_{\mathrm{BS}}\right)_{m,k_{\tau}^\star}$. In OpenISAC, the short-time Fourier transform (STFT) is employed to compute the spectrogram of $r_{\mathrm{BS}}[m]$. The STFT of $r_{\mathrm{BS}}[m]$ on windowed frames of length $M_w$ with hop size $M_H$ and analysis window function $w_{\mathrm{md}}$ is
\begin{equation}
\label{eq:md_stft}
\begin{aligned}
&\left(\boldsymbol{G}\right)_{m,k_f}
=\sum_{\ell=0}^{M_w-1} r_{\mathrm{BS}}\!\left[mM_H+\ell\right]\; w_\mathrm{md}[\ell]\;
e^{-j2\pi \frac{k_f\,\ell}{M_{\mathrm{md}}}},
\\
&k_f\in\{0,1,\ldots,M_{\mathrm{md}}-1\},
\end{aligned}
\end{equation}
where $M_{\mathrm{md}}\!\ge\!M_w$ is the DFT size, $m$ is the frame index, and $\boldsymbol{G}$ is the STFT matrix. The spectrogram is then calculated as
\begin{equation}
\label{eq:md_spectrogram}
\left( \mathrm{SPT} \right) _{m,k_f}=\frac{1}{M_w}\left| \left( \boldsymbol{G} \right) _{m,k_f} \right|^2.
\end{equation}
We display $\mathrm{SPT}$ in two-sided form (after an FFT shift) so that the zero-Doppler bin is centered.
\subsection{UE Communication Reception} \label{subsec:UE_ComRx}

\begin{figure}[!htbp]
\centering
\includegraphics[width=0.48\textwidth]{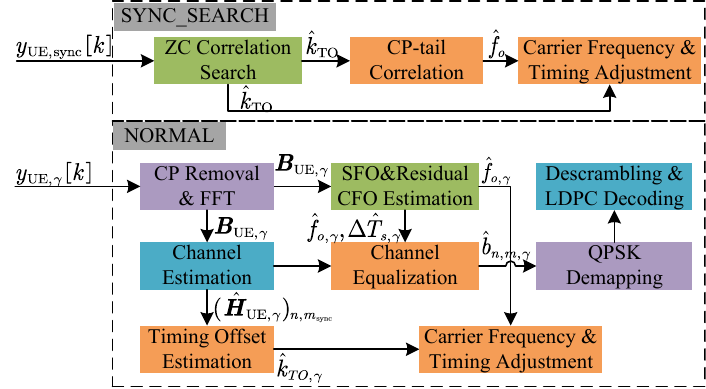}
\caption{Signal processing procedure for UE communication reception.}
\label{fig:flowgraph_ue}
\end{figure}

To enable robust real-time operation, the UE receiver logic is organized into a finite state machine with two states: \textit{SYNC\_SEARCH} and \textit{NORMAL}, as illustrated in Fig.~\ref{fig:flowgraph_ue}.
Reception begins in the \textit{SYNC\_SEARCH} state, which is responsible for detecting frame boundaries and estimating coarse carrier frequency offsets.
Once the initial timing and frequency offsets are compensated, the receiver transitions to the \textit{NORMAL} state.
In this state, the UE performs continuous OFDM demodulation, fine synchronization tracking, and payload decoding.
Throughout the process, the synchronization status is monitored; if the timing or frequency offsets exceed the reliable range or the signal strength falls below a threshold, the UE declares a loss of lock and reverts to the \textit{SYNC\_SEARCH} state.

\subsubsection{SYNC\_SEARCH State}
The UE operates in a block-by-block fashion: in each iteration, a block of complex baseband samples is fetched from the USRP and processed. Let $N_s \triangleq N+N_{\mathrm{CP}}$ denote the number of samples in one OFDM symbol. In the \textit{SYNC\_SEARCH} state, the UE acquires $2MN_s$ samples per iteration and then searches for the synchronization fields to obtain initial timing and frequency-offset estimates. The block length $2M N_s$ is chosen to ensure that at least one complete synchronization symbol is present in the block.
The discrete-time synchronization symbol $s_{\mathrm{ZC}}[k]$ is generated via IFFT of the frequency-domain ZC sequence, which preserves the ideal autocorrelation properties~\cite{popovic2010efficient}. In the compact one-ZC-symbol configuration, the UE estimates the initial timing offset by performing a sliding correlation between the received block and the local ZC reference $s_{\mathrm{ZC}}[k]$. To enhance robustness against amplitude fluctuations, the normalized correlation energy is used as the detection metric:
\begin{equation}
\Lambda_{\mathrm{ZC}}[u]
=\frac{\left|\sum_{k=0}^{N_s-1}y_{\mathrm{UE}}[u+k]s_{\mathrm{ZC}}^{*}[k]\right|^2}
{\left(\sum_{k=0}^{N_s-1}|y_{\mathrm{UE}}[u+k]|^2\right)
\left(\sum_{k=0}^{N_s-1}|s_{\mathrm{ZC}}[k]|^2\right)}.
\label{eq:zc_sync_metric}
\end{equation}
The peak location $\hat{k}_{\mathrm{peak}}$ is then obtained by finding the global maximum of $\Lambda_{\mathrm{ZC}}[u]$. Let $\mathcal{U}$ denote the search window. The implementation compares the peak-to-average correlation ratio
\begin{equation}
\rho_{\mathrm{ZC}}\triangleq
\frac{\Lambda_{\mathrm{ZC}}[\hat{k}_{\mathrm{peak}}]}
{\frac{1}{|\mathcal{U}|}\sum_{u\in\mathcal{U}}\Lambda_{\mathrm{ZC}}[u]},
\end{equation}
with a threshold $\rho_{\mathrm{th}}$. The threshold is selected above the noise-only correlation floor to avoid false synchronization, while remaining sufficiently low to accept the synchronization peak at relatively low SNR. In the reported experiments, $\rho_{\mathrm{th}}=100$. If $\rho_{\mathrm{ZC}}>\rho_{\mathrm{th}}$, the timing estimate is given by
\begin{equation}
\hat{k}_{\mathrm{TO}}=\hat{k}_{\mathrm{peak}}-m_{\mathrm{sync}}N_s-N_{\mathrm{lag}}.
\label{eq:initial_timing_est}
\end{equation}
Note that $\hat{k}_{\mathrm{TO}}$ corresponds to the aggregate delay composed of the initial timing offset $\tau_d$  and the additional propagation delay $\tau_1$ of the strongest path, i.e., $\hat{k}_{\mathrm{TO}}\approx B\left( \tau_1+\tau_d \right)-N_{\mathrm{lag}}$. The timing offset is calculated relative to the $N_{\mathrm{lag}}$th sample of the frame. This is to accommodate multipath components that arrive earlier than the strongest path (i.e., $\tau_l<\tau_1$), so that the subsequent corrections would not cause ISI. Therefore, $N_{\mathrm{lag}}$ is selected to be smaller than $N_{\mathrm{CP}}$ and large enough to leave room before the detected strongest path for earlier-arriving multipath components. In the reported experiments, $N_{\mathrm{lag}}$ is set to $20$ samples, while $N_{\mathrm{CP}}=128$. When $\hat{k}_{\mathrm{TO}}>0$, the timing offset is later compensated by fetching $\hat{k}_{\mathrm{TO}}$ additional samples at the beginning of the next processing block and discarding them before further processing. When $\hat{k}_{\mathrm{TO}}<0$, the stream is effectively ahead by $|\hat{k}_{\mathrm{TO}}|$ samples. In this case we acquire $|\hat{k}_{\mathrm{TO}}|$ fewer samples and prepend the same number of zeros to the front so that the FFT windows remain properly aligned with the OFDM symbols. After compensation, subsequent blocks are naturally aligned to the frame boundaries.

For operation with larger initial CFO, the optional second ZC synchronization symbol enables a Schmidl-Cox-type acquisition metric~\cite{schmidl1997robust}. For a trial start index $u$, the receiver computes
\begin{equation}
\begin{aligned}
P_{\mathrm{SC}}[u]
&=\sum_{k=0}^{N-1}y_{\mathrm{UE}}^{*}[u+N_{\mathrm{CP}}+k]
y_{\mathrm{UE}}[u+N_s+N_{\mathrm{CP}}+k],\\
R_{\mathrm{SC}}[u]
&=\sum_{k=0}^{N-1}\left|y_{\mathrm{UE}}[u+N_s+N_{\mathrm{CP}}+k]\right|^2,\\
\Lambda_{\mathrm{SC}}[u]&=\frac{|P_{\mathrm{SC}}[u]|^2}{R_{\mathrm{SC}}^2[u]}.
\end{aligned}
\end{equation}
The start index and modulo CFO can then be estimated as
\begin{equation}
\begin{aligned}
&\hat{u}= \arg\max_u \Lambda_{\mathrm{SC}}[u],\\
&\hat f_{o,\mathrm{mod}}= \frac{\angle P_{\mathrm{SC}}[\hat{u}]}{2\pi N_sT_s}.
\end{aligned}
\end{equation}
The modulo CFO estimate has an unambiguous range of $\pm1/(2N_sT_s)$. For a configured CFO search range, the receiver evaluates candidate CFOs separated by $1/(N_sT_s)$; each candidate is derotated and tested by the local ZC-correlation metric in \eqref{eq:zc_sync_metric}, and the strongest peak is selected. The main ZC symbol is then used for the fine STO estimate in \eqref{eq:initial_timing_est}.

The optional CFO training field provides another ambiguity-resolution reference. Its useful part has repetition period $N_{\mathrm{CFO}}$, so the CFO can be estimated from the phase difference between repeated segments, with an unambiguous range of $\pm1/(2N_{\mathrm{CFO}}T_s)$. This estimate is used only to choose among the candidate CFOs. After candidate selection, CP-tail correlation~\cite{vandeBeek1997ML} refines the frequency estimate. If the second-ZC path is not used or not detected, the receiver instead obtains a modulo CFO estimate directly from CP-tail correlation over the available OFDM symbols, resolves the ambiguity using either the local ZC-correlation metric or the optional CFO-training estimate, and then compensates the resulting frequency estimate by digital frequency retuning or reference-clock adjustment.

\subsubsection{NORMAL State}
After the initial frequency and timing offset estimation, the UE transitions to the \textit{NORMAL} state. In this state, each iteration processes a block of $M N_s$ samples, i.e., the processing block length is exactly one frame. Without loss of generality, assume that the $\gamma$th received block aligns with
the $\gamma$th transmitted frame. By applying the timing and frequency adjustments described above to the received samples in \eqref{eq:bi_rx_samp}, the receiver forms frame-aligned blocks of length $MN_s$ at each iteration. The received signal in the $\gamma$th block can then be written as
\begin{equation}
\label{eq:y_ue_block}
y_{\mathrm{UE},\gamma}[k]=\!\sum_{l=1}^L{y_{l,\gamma}\left[ k \right]}+z_{\mathrm{UE},\gamma}\left[ k \right],\ k=0,\ldots,MN_s-1,
\end{equation}
where the signal from the $l$th path is
\begin{equation}
\begin{aligned}
y_{l,\gamma}\left[ k \right] \!=\!\alpha _ls_{\gamma}(kT_s\!-\!\tau _l\!-\!\bar{\tau}_{d,\gamma,k}\!)e^{j2\pi (f_{D,l}\!+\!\Delta \bar{f}_{c,\gamma})kT_s},
\end{aligned}
\end{equation}
where $\bar{\tau}_{d,\gamma,k}$ and $\!\Delta \bar{f}_{c,\gamma}$ can be expressed as
\begin{equation}
\begin{aligned}
&\bar{\tau}_{d,\gamma ,k}\!=\!\tau _d\!\!+\!k\Delta T_{s,\gamma}\!+\!\sum_{\gamma ^{\prime}=0}^{\gamma -1}{\left( MN_s\Delta T_{s,\gamma ^{\prime}}-\!\hat{k}_{\mathrm{TO},\gamma ^{\prime}}T_s \right) ,}
\\
&\Delta \bar{f}_{c,\gamma}=\Delta f_c-\sum_{\gamma ^{\prime}=0}^{\gamma -1}{\hat{f}_{o,\gamma ^{\prime}}},
\end{aligned}
\label{eq:CFO_STO_def}
\end{equation}
with $\hat{k}_{\mathrm{TO},\gamma}T_s$ and $\hat{f}_{o,\gamma}$ denoting the timing and frequency compensations applied from the $\gamma$th block. $\Delta T_{s,\gamma}$ is the SIO during the $\gamma$th block. Note that if the frequency offset is corrected by adjusting the reference clock, the SIO $\Delta T_{s,\gamma}$ is compensated as well, because both the local oscillator (LO) and the sampling clock are locked to the same reference. In contrast, when a digital frequency retune is applied, the sampling frequency is unchanged and the SIO remains uncorrected, i.e., $\Delta T_{s,\gamma}=\Delta T_s$. Assuming that the maximum relative delay $\tau _{\max,\gamma}={\max}_l\left\{ \tau_{l}
+\bar{\tau}_{d,\gamma,k} \right\}$ does not exceed the CP duration $T_\mathrm{CP}$, and that the maximum relative Doppler $f_{D,\max,\gamma}={\max}_l\left\{ |f_{D,l}+\Delta \bar{f}_{c,\gamma} | \right\}$ is smaller than $\Delta f/10$, then after CP removal, the received signal of the $\gamma$th block can be rearranged into $N\times M$ receive matrix $ \boldsymbol{B}_{\mathrm{UE},\gamma}$ using FFT,
\begin{equation}
\begin{aligned}
\left( \!\boldsymbol{B}_{\mathrm{UE},\gamma}\! \right) _{n,m}\!=\!b_{n,m,\gamma}\!\left( \boldsymbol{H}_{\mathrm{UE},\gamma} \right) _{n,m}+\!\left( \boldsymbol{Z}_{\mathrm{UE},\gamma} \right) _{n,m},
\end{aligned}
\label{eq:ue_rx_grid}
\end{equation}
where $\boldsymbol{H}_{\mathrm{UE},\gamma}$ is the channel matrix for the $\gamma$th frame expressed as
\begin{equation}
\begin{aligned}
\left( \boldsymbol{H}_{\mathrm{UE},\gamma} \right) _{n,m}\!=\!\sum_{l=1}^L{\alpha _le^{j2\pi ((f_{D,l}+\Delta \bar{f}_{c,\gamma})mT_O\!-\!n\Delta f(\tau _l\!+\!\bar{\tau}_{d,\gamma,mN_s}))}}.
\end{aligned}
\label{eq:ue_rx_channel}
\end{equation}
For communication, assuming the Doppler spread is negligible across $M$ OFDM symbols, the channel matrix can be approximated by,
\begin{equation}
\begin{aligned}
\left( \boldsymbol{H}_{\mathrm{UE},\gamma} \right) _{n,m}\!\approx \!e^{j2\pi(f_{D,1}\!+\!\Delta \bar{f}_{c,\gamma})mT_O}\!\sum_{l=1}^L{\alpha _le^{\!-\!j2\pi \!n\Delta f(\tau _l\!+ \bar{\tau}_{d,\!\gamma,\!mN_s}\!)}\!}.
\end{aligned}
\label{eq:ue_rx_channel_approx}
\end{equation}
We then estimate the channel of the $m_{\mathrm{sync}}$th OFDM symbol using the full-band ZC symbol as
\begin{equation}
\begin{aligned}
&( \hat{\boldsymbol{H}}_{\mathrm{UE},\gamma} ) _{n,m_{\mathrm{sync}}}=\frac{\left( \boldsymbol{B}_{\mathrm{UE},\gamma} \right) _{n,m_{\mathrm{sync}}}}{z_n}.
\end{aligned}
\label{eq:ue_rx_channel_esti}
\end{equation}
Based on the channel estimate, we first construct the complex delay spectrum, defined as  
\begin{equation}
\begin{aligned}
\tilde{p}_{\mathrm{delay},\gamma}\left[ k \right] 
&\triangleq \frac{1}{\sqrt{N}} \sum_{n=0}^{N-1}{(\hat{\boldsymbol{H}}_{\mathrm{UE},\gamma})_{n,m_{\mathrm{sync}}} e^{\frac{j2\pi nk}{N}}},
\end{aligned}
\label{eq:ue_rx_channel_delay_complex}
\end{equation}
where $k=0,\ldots,N-1$. The corresponding power delay spectrum is then given by  
\begin{equation}
\begin{aligned}
p_{\mathrm{delay},\gamma}\left[ k \right] 
&= \left| \tilde{p}_{\mathrm{delay},\gamma}\left[ k \right] \right|^2.
\end{aligned}
\label{eq:ue_rx_channel_delay}
\end{equation}
Let $k_{\max,\gamma}=\underset{k=0,...,N-1}{\mathrm{argmax}}\left( p_{\mathrm{delay},\gamma}\left[ k \right] \right)$ denote the index of the maximum value in \eqref{eq:ue_rx_channel_delay}. The timing offset relative to the $N_{\mathrm{lag}}$th sample can then be estimated as
\begin{equation}
\begin{aligned}
\hat{k}_{TO,\gamma}=\begin{cases}
	k_{\max,\gamma}-N_{\mathrm{lag},\gamma}, &\mathrm{if}\ k_{\max,\gamma}\le N/2,\\
	k_{\max,\gamma}-N-N_{\mathrm{lag},\gamma}, &\mathrm{if}\ k_{\max,\gamma}>N/2,\\
\end{cases}
\end{aligned}
\label{eq:ue_rx_channel_offset}
\end{equation}
which is subsequently used for timing correction.
Note that the timing-offset drift caused by SIO within one frame can also be estimated from the CFO. When the sampling clock and carrier LO are derived from the same reference, let $\eta_{\mathrm{ppm}}$ denote the relative clock deviation in ppm and $\eta=\eta_{\mathrm{ppm}}\times10^{-6}$. Then the CFO and SFO satisfy
\begin{equation}
\begin{aligned}
\Delta f_c &= \eta f_c,\\
\Delta f_s &= \eta B .
\end{aligned}
\end{equation}
Since $T_s=1/B$ and $T_{s,\mathrm{UE}}=T_s-\Delta T_s$, we have
\begin{equation}
\begin{aligned}
T_{s,\mathrm{UE}}
&=\frac{1}{B+\Delta f_s}
=\frac{T_s}{1+\eta}
\approx T_s(1-\eta),\\
\Delta T_s
&=T_s-T_{s,\mathrm{UE}}
=\frac{\eta}{B(1+\eta)}
\approx \frac{\eta}{B}
=\frac{\Delta f_c}{f_c B}.
\end{aligned}
\end{equation}
Therefore, the expected timing-offset drift over one frame is
\begin{equation}
\Delta k_{\mathrm{SIO}}^{\mathrm{frame}}
\approx M N_s B\Delta T_s
\approx M N_s\frac{\Delta f_c}{f_c}.
\end{equation}
Using this prediction, OpenISAC provides an optional \textit{predictive\_delay} option that pre-compensates the timing-offset and estimates only the residual timing-offset during alignment/tracking. When the SIO is relatively large, this option helps keep the dominant delay peak inside the tracking window and reduces the risk of losing lock.
Next, we estimate the frequency offset and SIO using the pilot subcarriers. Recall from \eqref{eq:bnm_mapping} that $b_{n,m,\gamma}=z_n, \forall  n \in \mathcal{P}$. Thus, we can calculate the cross-symbol autocorrelation as
\begin{equation}
\begin{aligned}
&\left( \!\boldsymbol{R}_{\mathrm{UE},\gamma}\! \right) _{n,m}\!=\left( \!\boldsymbol{B}_{\mathrm{UE},\gamma}^{*}\! \right) _{n,m}\left( \!\boldsymbol{B}_{\mathrm{UE},\gamma}\! \right) _{n,m+1}
\\
&=\frac{P_{\mathrm{Tx}}}{N}\!\left( \boldsymbol{H}_{\mathrm{UE},\gamma}^{*} \right) _{n,m}\left( \boldsymbol{H}_{\mathrm{UE},\gamma} \right) _{n,m+1}\!+\!( \mathring{\boldsymbol{Z}}_{\mathrm{UE},\gamma} ) _{n,m},
\\
&\forall n\in \mathcal{P},\ m=0,\ldots,M-2, 
\end{aligned}
\label{eq:ue_rx_channel_delay_corr}
\end{equation}
where the noise term can be written as $(\mathring{\boldsymbol{Z}}_{\mathrm{UE},\gamma})_{n,m}=\!\left( \boldsymbol{Z}_{\mathrm{UE},\gamma}^{*} \right) _{n,m}\left( \boldsymbol{H}_{\mathrm{UE},\gamma} \right) _{n,m+1}+\!\left( \boldsymbol{Z}_{\mathrm{UE},\gamma} \right) _{n,m+1}\left( \boldsymbol{H}_{\mathrm{UE},\gamma}^{*} \right) _{n,m}+\left( \boldsymbol{Z}_{\mathrm{UE},\gamma}^{*} \right) _{n,m}\left( \boldsymbol{Z}_{\mathrm{UE},\gamma} \right) _{n,m+1}$. By substituting \eqref{eq:CFO_STO_def} and \eqref{eq:ue_rx_channel_approx}, $\left( \boldsymbol{H}_{\mathrm{UE},\gamma}^{*} \right) _{n,m}\left( \boldsymbol{H}_{\mathrm{UE},\gamma} \right) _{n,m+1}$ can be further written as \eqref{eq:ue_rx_channel_corss}.
\begin{figure*}[!t]
\centering
\begin{equation}
\begin{aligned}
&\left( \boldsymbol{H}_{\mathrm{UE},\gamma}^{*} \right)_{n,m}
\left( \boldsymbol{H}_{\mathrm{UE},\gamma} \right)_{n,m+1} \approx 
\mathrm e^{j2\pi\big[(f_{D,1}+\Delta \bar f_{c,\gamma})T_O
- n\Delta f N_s \Delta T_{s,\gamma}\big]}
\Bigg[
\underbrace{\sum_{l=1}^L |\alpha_l|^2}_{\text{same-path term}}
+
\underbrace{\sum_{\substack{l,l'=1\\ l'\neq l}}^{L}
\alpha_l^{*}\alpha_{l'}
\mathrm e^{j2\pi n\Delta f(\tau_l-\tau_{l'})}
}_{\text{multi-path cross term}}
\Bigg].
\end{aligned}
\label{eq:ue_rx_channel_corss}
\end{equation}
\end{figure*}
The multi-path cross term is relatively small compared to the dominant same-path component, because it results from non-coherent accumulation over different paths with varying phases, and can therefore be treated as a perturbation. Since \eqref{eq:ue_rx_channel_corss} is independent of the symbol index $m$, we average the autocorrelation across OFDM symbols to suppress the noise term, i.e.,
\begin{equation}
\begin{aligned}
\bar{R}_{\mathrm{UE},\gamma}[n]
&\triangleq \frac{1}{M-1}\sum_{m=0}^{M-2}
\left( \boldsymbol{R}_{\mathrm{UE},\gamma} \right)_{n,m},
\quad \forall n\in \mathcal{P}.
\end{aligned}
\label{eq:ue_rx_channel_delay_corr_avg}
\end{equation}
Next, we compute the phase of \eqref{eq:ue_rx_channel_delay_corr_avg} as
\begin{equation}
\begin{aligned}
\varphi _{\mathrm{UE},\gamma}[n]
&\triangleq \arg\!\left( \bar{R}_{\mathrm{UE},\gamma}[n] \right) 
\\
&\approx 2\pi \big( f_{o,\gamma}T_O - n\Delta f\,N_s\Delta T_{s,\gamma} \big),
\end{aligned}
\label{eq:ue_rx_channel_delay_corr_phase}
\end{equation}
where $f_{o,\gamma}=f_{D,1}+\Delta \bar{f}_{c,\gamma}$ denotes the effective frequency offset during the $\gamma$th frame. After phase unwrapping over the pilot subcarriers, \eqref{eq:ue_rx_channel_delay_corr_phase} becomes a linear function of $n$, so weighted linear regression with weights $\big|\bar{R}_{\mathrm{UE},\gamma}[n]\big|^2$ yields
\begin{equation}
\hat{\boldsymbol{\theta}}_\gamma
\triangleq
\begin{bmatrix}
\hat{f}_{o,\gamma} \\
\Delta \hat{T}_{s,\gamma}
\end{bmatrix}.
\label{eq:ue_wls_fo_dTs}
\end{equation}
The estimate $\hat{f}_{o,\gamma}$ is used for carrier-frequency compensation. Then, the channel estimates over the entire $\gamma$th frame are obtained by propagating the estimate at the synchronization symbol $m_{\mathrm{sync}}$ as
\begin{equation}
\begin{aligned}
&(\hat{\boldsymbol{H}}_{\mathrm{UE},\gamma})_{n,m}=(\hat{\boldsymbol{H}}_{\mathrm{UE},\gamma})_{n,m_{\mathrm{sync}}}\exp \bigl( j2\pi ( m-m_{\mathrm{sync}} ) \cdot \bigr.
\\
&\left. ( \hat{f}_{o,\gamma}T_O-n\Delta fN_s\Delta \hat{T}_{s,\gamma} ) \right), \forall\, m,n.
\end{aligned}
\label{eq:channel_estimation_complete}
\end{equation}
Based on the above channel estimates, the detected data symbols are obtained via one-tap frequency-domain equalization as
\begin{equation}
\begin{aligned}
\hat{b}_{n,m,\gamma}=\frac{\left( \boldsymbol{B}_{\mathrm{UE},\gamma} \right) _{n,m}}{(\hat{\boldsymbol{H}}_{\mathrm{UE},\gamma})_{n,m}},\forall m\neq m_{\mathrm{sync}},\, \forall n \notin \mathcal{P}.
\end{aligned}
\label{eq:channel_equalization}
\end{equation}
The equalized data symbols $\hat{b}_{n,m,\gamma}$ are then used to compute the LLRs for LDPC soft decoding, which is omitted here for brevity.

\subsection{UE Bistatic Sensing} \label{subsec:UE_Sensing}
The signal processing for bistatic sensing largely follows that of the monostatic case discussed in Sec.~\ref{subsec:monosensing}, but with two key differences. First, the modulation symbols are not known a priori at the receiver and must therefore be reconstructed. Second, transmitter-receiver synchronization is relatively straightforward for monostatic sensing, since the transmitter and receiver are co-located, whereas in bistatic sensing a wired synchronization link is often unavailable, so OTA synchronization is required instead.

\subsubsection{Modulation Symbol Reconstruction} \label{subsec:UE_Symbol_Reconst}
For bistatic sensing, the data symbols on $n\in\mathcal{D}$ are reconstructed from the equalized symbols $\hat{b}_{n,m,\gamma}$ by hard decision over the QPSK constellation,
\begin{equation}
\tilde{b}_{n,m,\gamma}
= \frac{1}{\sqrt{2}}\big(
\operatorname{sgn}(\mathrm{Re}\{\hat{b}_{n,m,\gamma}\})
+ j\operatorname{sgn}(\mathrm{Im}\{\hat{b}_{n,m,\gamma}\})
\big).
\label{eq:ue_qpsk_reconst_impl}
\end{equation}
For the synchronization symbol and pilot subcarriers, the transmitted symbols are already known as $z_n$ and are used directly.

\subsubsection{OTA Synchronization} \label{subsec:UE_Sync}
Note that the timing corrections for communication, as described in Sec.~\ref{subsec:UE_ComRx} are insufficient for bistatic sensing. For communication, as long as the maximum relative delay $\tau _{\max,\gamma}={\max}_l\left\{ \tau_{l}+\bar{\tau}_{d,\gamma,k} \right\}$ does not exceed the CP duration $T_\mathrm{CP}$, ISI is avoided and the system operates properly. In that case, sub-sample timing drift mainly appears as a phase term in the channel estimate and is usually tolerable for communication reception. Consequently, the receiver only needs to apply an explicit timing update when the accumulated drift becomes large enough for the dominant delay peak to cross into an adjacent delay bin, i.e., approximately one sampling interval $T_s$. For bistatic sensing, however, delay is itself the quantity of interest. Reusing these sporadic integer-sample corrections therefore creates a staircase delay trajectory, which appears as abrupt jumps in the delay-Doppler or micro-Doppler outputs. In this subsection, we therefore propose a low-complexity OTA bistatic synchronization scheme that uses the line-of-sight (LoS) communication link as a timing reference, while incorporating additional practical considerations to enable robust real-time implementation.

First, we refine the timing-offset estimate to obtain a fractional result. A conventional approach is to perform interpolation by zero-padding in the subcarrier domain prior to the delay-domain FFT. However, accurate estimation typically requires a high interpolation order, which significantly increases computational complexity. To enable real-time processing, we instead adopt Quinn’s fractional frequency-estimation algorithm \cite{295186}. Denote the fractional timing offset by $\delta_{\tau} \triangleq \tau_{o,\gamma}N \Delta f - k_{\max,\gamma}$, where $\tau_{o,\gamma} = \tau_1 + \bar{\tau}_{d,\gamma,N_sm_{\mathrm{sync}}}$ is the overall timing offset. Define  
\begin{equation}
r_p\left[ k \right] \triangleq \frac{\tilde{p}_{\mathrm{delay},\gamma}\left[ k_{\max ,\gamma}+k \right]}{\tilde{p}_{\mathrm{delay},\gamma}\left[ k_{\max ,\gamma} \right]},\quad k\in\{-1,1\}.
\label{eq:ratio_fractional}
\end{equation}
According to \cite{295186}, $r_p\left[ 1 \right] \approx \frac{\delta_{\tau}}{\delta_{\tau}-1}$ and $r_p\left[ -1 \right] \approx \frac{\delta_{\tau}}{\delta_{\tau}+1}$. We thus obtain two candidate estimates of the fractional timing offset,
\begin{equation}
\hat{\delta}_{\tau,1}=\frac{r_p\left[ 1 \right]}{r_p\left[ 1 \right] -1},\quad 
\hat{\delta}_{\tau,-1}=\frac{r_p\left[ -1 \right]}{1-r_p\left[ -1 \right]}.
\label{eq:fractional_delay_two_estimate}
\end{equation}
When $\delta_{\tau} > 0$, $\hat{\delta}_{\tau,1}$ is generally more accurate. Otherwise, $\hat{\delta}_{\tau,-1}$ is more accurate. Since $\delta_{\tau}$ is unknown in practice, its sign is inferred from the two candidate estimates. Following the selection rule in \cite{295186}, if both candidates are positive, the offset is treated as positive and $\hat{\delta}_{\tau,1}$ is selected. Accordingly, the final estimate is selected as follows,
\begin{equation}
\hat{\delta}_{\tau}=
\begin{cases}
\hat{\delta}_{\tau,1}, & \text{if } \hat{\delta}_{\tau,-1} > 0 \ \text{and}\  \hat{\delta}_{\tau,1} > 0,\\[1mm]
\hat{\delta}_{\tau,-1}, & \text{otherwise}.
\end{cases}
\label{eq:fractional_delay_estimate}
\end{equation}
Let $k_{\tau,\gamma}$ denote the overall timing offset in samples. The corresponding timing offset in seconds is then estimated as
\begin{equation}
\hat{\tau}_{o,\gamma}
=
\frac{\hat{k}_{\tau ,\gamma}}{B}
=
\frac{\hat{\delta}_{\tau}+k_{\max ,\gamma}}{B}.
\label{eq:overall_delay_estimate}
\end{equation}
However, since the fractional estimates obtained from Quinn’s algorithm are not sufficiently accurate and are contaminated by noise, we need to further refine the timing-offset estimate. From \eqref{eq:CFO_STO_def}, we observe that the timing-offset drift is generally dominated by the SIO and thus follows an approximately linear trend. In the next step, we exploit this property to further improve the timing-offset estimation.

Note that $\hat{k}_{\mathrm{TO},\gamma}$ denotes the timing offset compensation applied from the $\gamma$th frame. We partition the frames into non-overlapping SIO estimation windows of length $\varGamma_W$ frames, and denote by $\gamma_w$ the starting frame index of the $w$th window. Within this window, we define  
\begin{equation}
A_{\gamma_w+\ell} \triangleq \sum_{i=0}^{\ell-1} \hat{k}_{\mathrm{TO},\gamma_w+i},\quad \ell = 0,\ldots,\varGamma_W-1,
\label{eq:alignment_def}
\end{equation}
so that $A_{\gamma_w+\ell}$ collects only the integer timing corrections applied inside the current window, up to frame $\gamma_w+\ell-1$.

To obtain a finer SIO estimate, we reconstruct a ``continuous'' delay trajectory by adding back these in-window corrections. Specifically, for each frame in the window we form
\begin{equation}
\tilde{k}_{\tau,\gamma_w+\ell} \triangleq \hat{k}_{\tau,\gamma_w+\ell} + A_{\gamma_w+\ell}.
\end{equation}
Over the $\varGamma_W$ frames in the window, we approximate the reconstructed delay samples by a linear model
\begin{equation}
\tilde{k}_{\tau,\gamma_w+\ell} \approx \epsilon_{\mathrm{SIO},w}\,\ell + k_{\tau,\gamma_w},
\end{equation}
and obtain the slope estimate $\hat{\epsilon}_{\mathrm{SIO},w}$ via standard least-squares linear regression. Here, $k_{\tau,\gamma_w}$ is the intercept of the linear model. Although its estimate $\hat{k}_{\tau,\gamma_w}$ is also obtained by the regression, the subsequent synchronization update uses only the slope estimate $\hat{\epsilon}_{\mathrm{SIO},w}$.
The SIO is determined by the sampling-clock mismatch at the transmitter and receiver. Given that the frequency drift of crystal oscillators is typically slow and small, especially when oven-controlled crystal oscillators (OCXOs) are used, the resulting delay evolution within a window of $\varGamma_W$ frames can be well approximated as linear, making the above model relatively accurate. The slope $\epsilon_{\mathrm{SIO},w}$ thus represents the SIO-induced delay drift per frame (in samples) in the $w$th window, and can be expressed as $\epsilon_{\mathrm{SIO},w} \triangleq M N_s \,\Delta {T}_{{as},w}B$,
where $\Delta {T}_{{as},w}$ is the average SIO over this window.

We then maintain a recursive estimate of the cumulative sensing timing-offset $\hat{k}^{\mathrm{sens}}_{\tau,\gamma}$ via
\begin{equation}
\hat{k}^{\mathrm{sens}}_{\tau,\gamma}
= 
\hat{k}^{\mathrm{sens}}_{\tau,\gamma-1}
+ \hat{\epsilon}_{\mathrm{SIO},w-1}
- \hat{k}_{\mathrm{TO},\gamma-1}
+ \mu_\gamma e_\gamma,
\label{eq:sensing_delay_update}
\end{equation}
where $w$ is the current SIO-estimation window index and $
e_\gamma \triangleq \hat{k}_{\tau,\gamma} - \hat{k}^{\mathrm{sens}}_{\tau,\gamma-1}$
denotes the tracking error between the instantaneous timing-offset estimate $\hat{k}_{\tau,\gamma}$ and its recursively predicted value $\hat{k}^{\mathrm{sens}}_{\tau,\gamma-1}$. Here, $\hat{k}_{\mathrm{TO},\gamma-1}$ is the integer timing correction applied from the previous frame, and $\mu_\gamma$ is a proportional gain. The term $\mu_\gamma e_\gamma$ introduces a first-order correction based on the current tracking error, which helps prevent the accumulation of errors caused by estimation inaccuracies and other unmodeled effects. When $|e_\gamma|$ remains small, $\mu_\gamma$ is kept at a small default value (e.g., $\mu_\gamma = 10^{-5}$) to avoid injecting noise. If $|e_\gamma|$ exceeds a predefined threshold (e.g., $0.1$ samples) for a sufficiently long period, $\mu_\gamma$ is increased (e.g., to $10^{-2}$) to accelerate convergence. The SIO estimator operates in real time: every $\varGamma_W$ frames, it updates $\hat{\epsilon}_{\mathrm{SIO},w}$ using a new, non-overlapping window, while \eqref{eq:sensing_delay_update} continuously refines $\hat{k}^{\mathrm{sens}}_{\tau,\gamma}$ using the latest available SIO estimate and the tracking-error feedback.

Next, the estimated timing offset $\hat{k}^{\mathrm{sens}}_{\tau,\gamma}$ and SIO $\Delta \hat{T}_{as,w}$ are used to compensate the sensing channel symbols. First, the OFDM channel symbols for bistatic sensing are obtained as
\begin{equation}
\left( \boldsymbol{F}_{\mathrm{UE},\gamma} \right) _{n,m}
= \frac{\left( \boldsymbol{B}_{\mathrm{UE},\gamma} \right) _{n,m}}{\tilde{b}_{n,m,\gamma}}.
\label{eq:bistatic_sensing_symbols}
\end{equation}
Then the SIO and timing-offset compensations are applied as
\begin{equation}
\left( \tilde{\boldsymbol{F}}_{\mathrm{UE},\gamma} \right) _{n,m}
=
\left( \boldsymbol{F}_{\mathrm{UE},\gamma} \right) _{n,m}
e^{j2\pi n\Delta f\left( \hat{k}_{\tau ,\gamma}^{\mathrm{sens}}+mN_s\Delta \hat{T}_{as,w-1} \right)}.
\label{eq:bistatic_sensing_symbols_comp}
\end{equation}
The compensated OFDM channel symbols $\tilde{\boldsymbol{F}}_{\mathrm{UE},\gamma}$ can then be used for various sensing tasks, in the same way as $\boldsymbol{F}_{\mathrm{BS},\gamma}$ in \eqref{eq:mono_rx_grid_channel}. Note that under this synchronization scheme, bistatic sensing yields the delays relative to the LoS path. Therefore, obtaining the absolute delays requires the known physical Tx-Rx separation (i.e., the LoS distance).

The above synchronization strategy is most reliable when the LoS path, or at least another stable dominant path, remains visible over time. In NLoS or rich-scattering environments, the strongest component may switch among multiple MPCs, in which case the reference path is no longer geometrically stable and the recovered delay trajectory may be biased or intermittently discontinuous. In such regimes, longer averaging, path-consistency checks, or external/network synchronization would be beneficial, and the receiver may need to fall back to re-acquisition through the synchronization state machine described in Sec.~\ref{subsec:UE_ComRx}.

\section{System Architecture}
\label{sec:sysarch}
\begin{figure}[!htbp]
\centering
\includegraphics[width=0.485\textwidth]{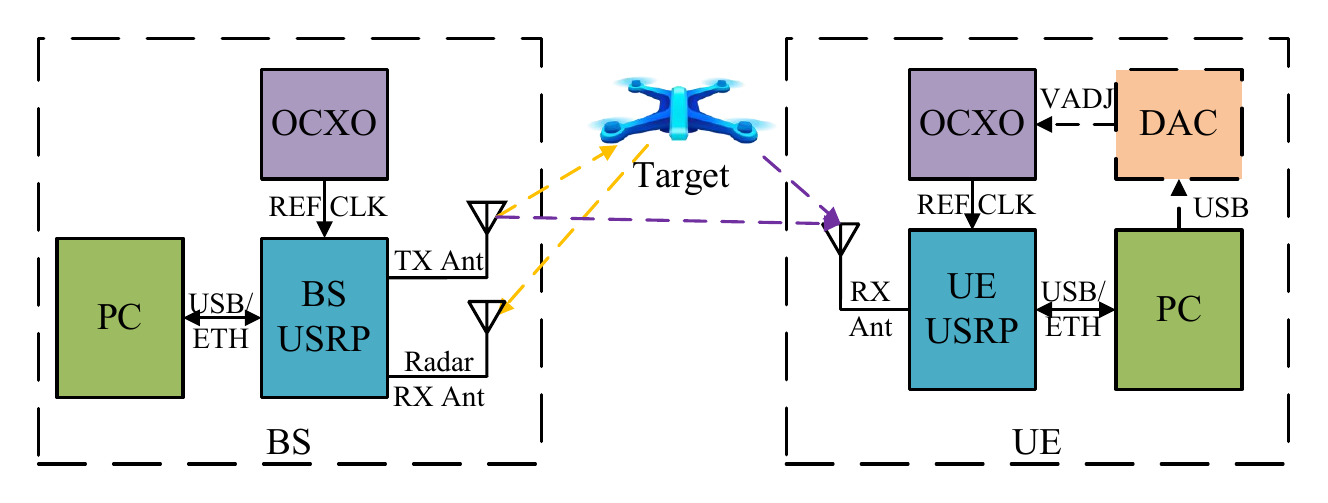}
\caption{System architecture of OpenISAC.}
\label{fig:sysarch}
\end{figure}
As shown in Fig.~\ref{fig:sysarch}, OpenISAC comprises a BS and a UE, both built around USRPs referenced by OCXOs. The BS host generates the OFDM-ISAC waveform, transmits it through a dedicated antenna, and receives radar echoes through a separate sensing antenna. The UE host receives the downlink waveform, performs real-time demodulation and bistatic sensing, and can optionally trim its OCXO through a USB DAC to reduce long-term CFO and SFO.

\subsection{Software Architecture of BS}

\begin{figure}[!htbp]
\centering
\includegraphics[width=0.485\textwidth]{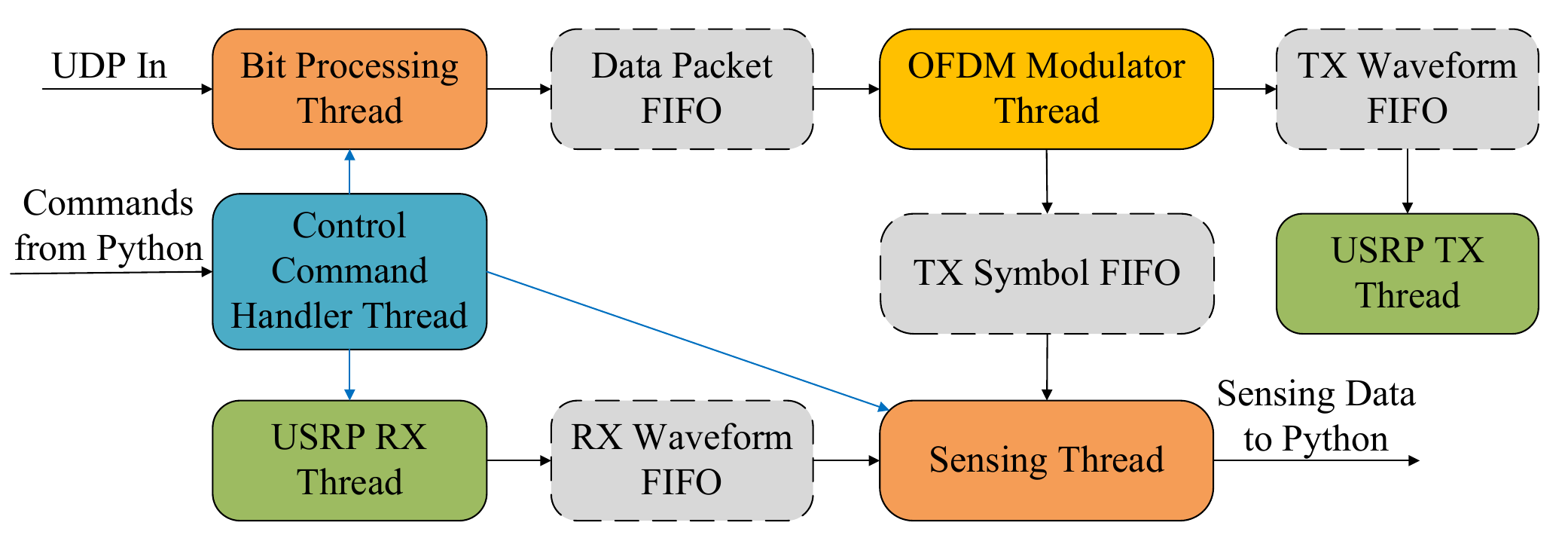}
\caption{Software architecture of BS.}
\label{fig:bs-sw-arch}
\end{figure}

As shown in Fig.~\ref{fig:bs-sw-arch}, the BS software is a multi-threaded FIFO pipeline. A bit-processing thread accepts UDP payloads, performs channel coding, and feeds the OFDM modulator, which generates transmit waveforms and stores the corresponding pre-IFFT symbols for sensing. Dedicated USRP-TX and USRP-RX threads handle timed transmission and continuous radar-stream capture.

A real-time sensing thread consumes paired RX frames and TX symbols, performs CP removal, FFT, and element-wise division, and by default computes range-Doppler maps. Its configurable stride parameter implements \eqref{eq:mono_rx_grid_channel_downsamp}, providing a practical load-resolution tradeoff, while a bypass mode streams the sensing channel to Python for custom processing. A lightweight control thread updates shared atomic variables so that the pipeline can be reconfigured without stopping real-time operation.

\subsection{Software Architecture of UE}
\begin{figure}[!htbp]
\centering
\includegraphics[width=0.485\textwidth]{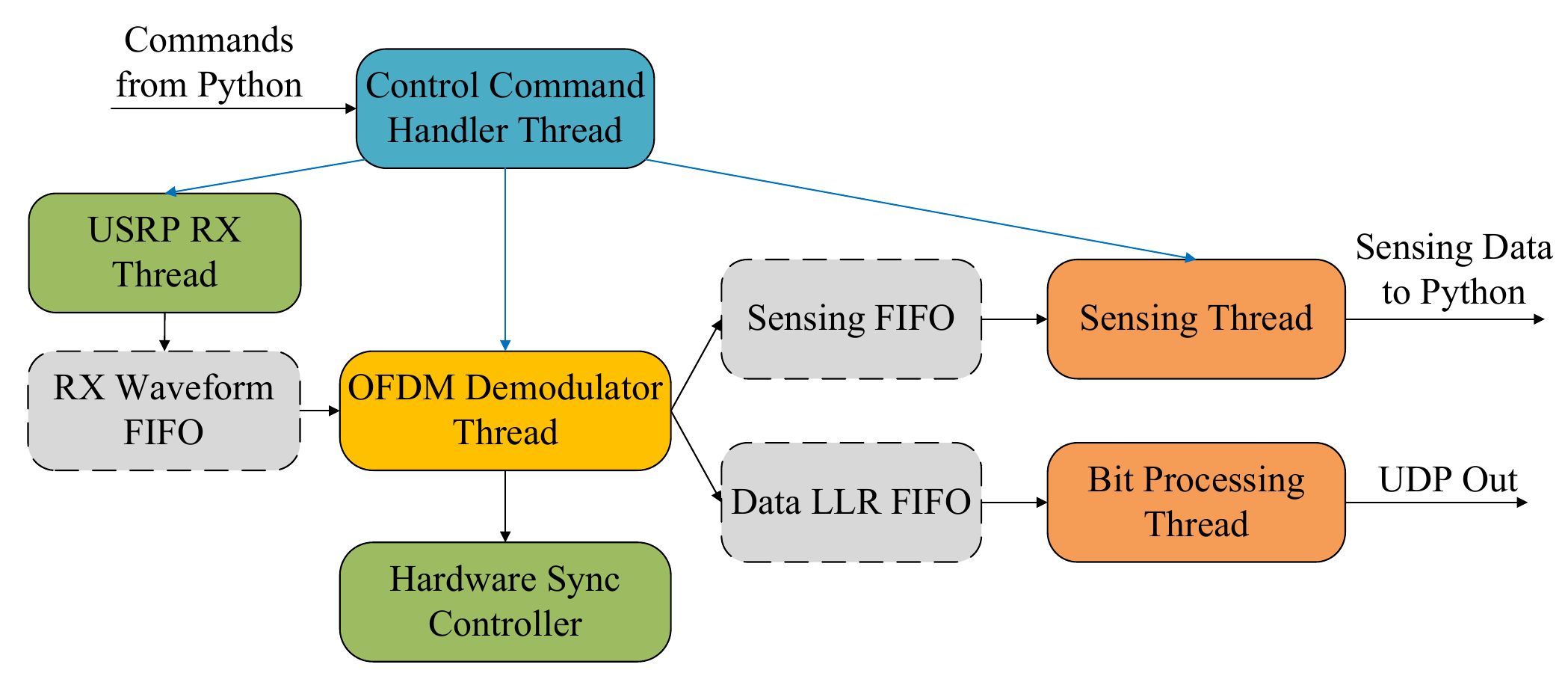}
\caption{Software architecture of UE.}
\label{fig:ue-sw-arch}
\end{figure}

As shown in Fig.~\ref{fig:ue-sw-arch}, the UE uses the same FIFO-based design philosophy. A USRP RX thread acquires the downlink stream and writes frame-aligned samples to the RX waveform FIFO. The OFDM demodulator thread executes the state machine in Sec.~\ref{subsec:UE_ComRx}, reconstructs the reference TX symbols, and emits both decoder inputs and \{RX, TX\} symbol pairs for bistatic sensing. A bit-processing thread performs LDPC decoding and forwards the recovered payload over UDP.

The sensing thread consumes the \{RX, TX\} symbol pairs, applies the demodulator-derived synchronization estimates, and computes bistatic sensing outputs in either normal delay-Doppler mode or a Python-facing bypass mode. In parallel, a lightweight control thread updates shared atomic variables, and the optional hardware sync controller can trim the UE OCXO to improve long-term coherence.

\section{Experimental Setup \& Results} \label{sec:exp_results}
\subsection{Experimental Setup}
\begin{figure}[!htbp]
\centering
\includegraphics[width=0.485\textwidth]{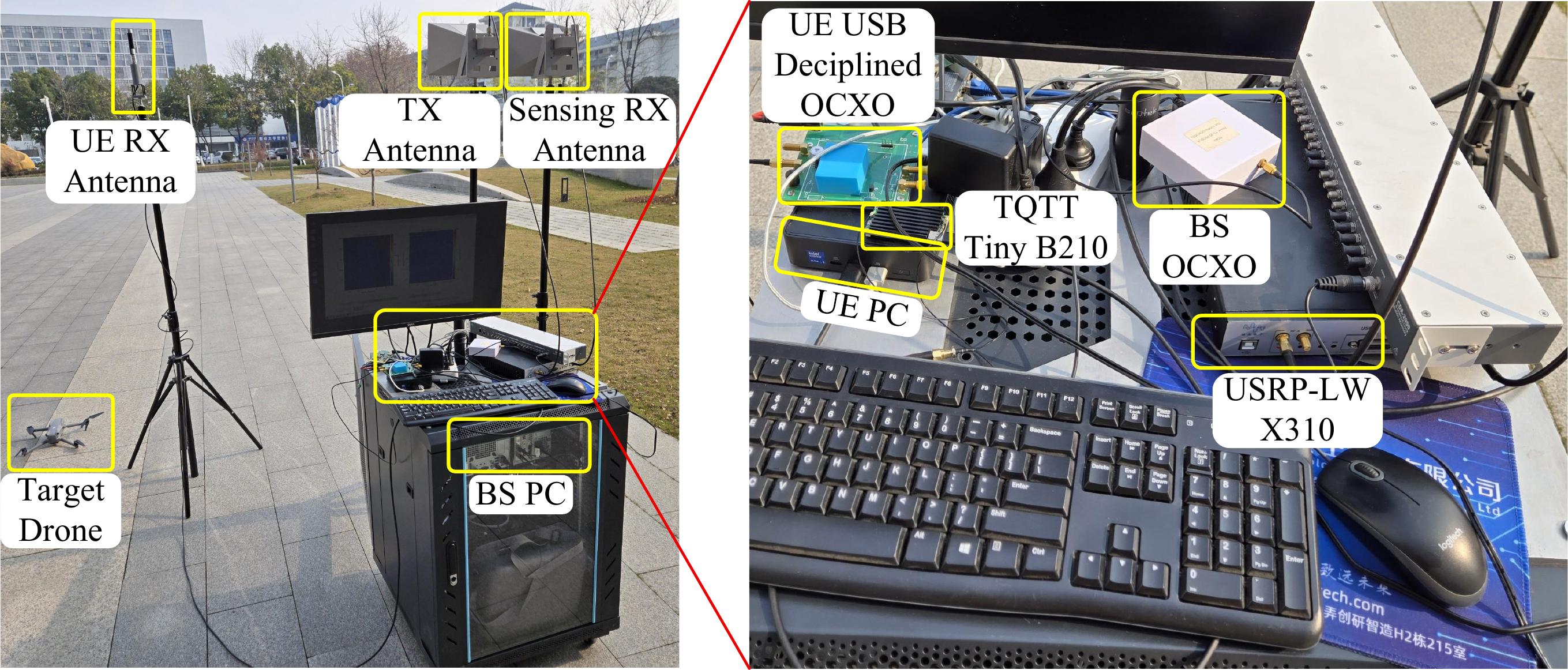}
\caption{Prototype implementation of the OpenISAC testbed, including the BS and UE nodes.}
\label{fig:expsetup}
\end{figure}
The prototype uses two hardware nodes, as shown in Fig.~\ref{fig:expsetup}. The BS employs a Luowave USRP-LW X310 connected to an Intel Core i7-10700 host over 10GbE, together with separate horn antennas for transmission and monostatic sensing reception, and an external OCXO (DAPU O23B-HCDD) as the reference clock. The UE uses a TQTT Tiny B210 connected to an Intel Core Ultra5 225H host over USB 3.0, a single omnidirectional receive antenna, and a Morion MV197 OCXO disciplined by a USB DAC. The over-the-air measurements are conducted in an outdoor open-area environment. The UE antenna is placed approximately 1.4~m to the front-left side of the BS transmit antenna. A DJI Mavic Air 3S UAV is used as the target in the monostatic micro-Doppler measurement in Fig.~\ref{fig:micro-Doppler_Mono} and the bistatic UAV sensing validation in Fig.~\ref{fig:bistatic_uav}.
For BS-side monostatic sensing, the separate transmit and sensing-receive horn antennas provide physical isolation between the transmit and receive paths. We also control the transmit power and receive gain so that the residual SI remains within the ADC dynamic range. Under this condition, the residual SI appears mainly as a zero-delay and zero-Doppler component and is suppressed together with static clutter by MTI. However, this simple SI handling is not a replacement for active analog SI cancellation. It cannot address strong SI that saturates or clips the ADC, as may occur in single-antenna ISAC operation or under large transmit power/receive gain. In addition, MTI suppresses static components and therefore cannot replace digital SI cancellation when static-target preservation is required. More advanced analog- and digital-domain SI cancellation techniques have been studied in~\cite{barneto2019fullduplex}.

Although this setup uses a high-end X310 at the BS, OpenISAC is hardware-agnostic and also supports lower-cost B200-series devices. Unless otherwise stated, the carrier frequency is $f_c=3.1$~GHz, the sample rate and analog bandwidth are both set to $B=50$~MHz, and a two-dimensional Hamming window is used in delay-Doppler processing. The detailed parameters are summarized in Table~\ref{tab:exp_parameters}.

\begin{table}[htbp]
\caption{System Parameters for Experimental Validation}
\label{tab:exp_parameters}
\centering
\renewcommand{\arraystretch}{1.2}
\begin{tabular}{|c|c|c|}
\hline
\textbf{Parameter} & \textbf{Notation} & \textbf{Value} \\ \hline
Carrier Frequency & $f_c$ & 3.1 GHz \\ \hline
TX Antenna Gain & $G_{\mathrm{TX}}$ & 16 dBi \\ \hline
Sensing RX Antenna Gain & $G_{\mathrm{RX,sens}}$ & 16 dBi \\ \hline
UE RX Antenna Gain & $G_{\mathrm{RX,UE}}$ & 3 dBi \\ \hline
TX Power & $P_{\mathrm{TX}}$ & 6.4 dBm \\ \hline
Sensing RX Gain & $G_{\mathrm{RX,sens}}^{\mathrm{RF}}$ & 10 dB \\ \hline
UE RX Gain & $G_{\mathrm{RX,UE}}^{\mathrm{RF}}$ & 30 dB \\ \hline
System Bandwidth \& Sample Rate & $B$ & 50 MHz \\ \hline
Number of Subcarriers & $N$ & 1024 \\ \hline
CP Length & $N_{\mathrm{CP}}$ & 128 \\ \hline
OFDM Symbols per Frame & $M$ & 100 \\ \hline
Sensing OFDM Symbols & $M_s$ & 100 \\ \hline
Symbol Stride & $M_D$ & 20 \\ \hline
STFT Window Length & $M_w$ & 256 \\ \hline
STFT Hop Size & $M_H$ & 64 \\ \hline
\end{tabular}
\end{table}
For USRPs, the sample rate is typically set slightly higher than the signal bandwidth. In our experiments, to use all available subcarriers and increase the sensing resolution, we set both the sample rate and analog bandwidth to 50~MHz.

\subsection{Experimental Results}
\subsubsection{\texorpdfstring{Real-Time Performance}{Real-Time Performance}}

To quantify the host-side real-time capability of OpenISAC, we ran benchmark sweeps for the baseband modulator, demodulator, and monostatic sensing pipeline on the same BS host platform. These sweeps characterize the software real-time margin and are therefore not restricted to the 50~MHz setting.

To evaluate the computational complexity and real-time performance of the baseband modulator and demodulator, we measured the host-side CPU usage and per-frame processing time at sample rates of 50, 100, and 200~MHz. The analog bandwidth is set equal to the sample rate, and the FFT size and CP length are increased proportionally, from 1024/128 to 2048/256 and 4096/512, so that the subcarrier spacing and frame duration remain unchanged.

\begin{figure}[!htbp]
\centering
\includegraphics[width=0.35\textwidth]{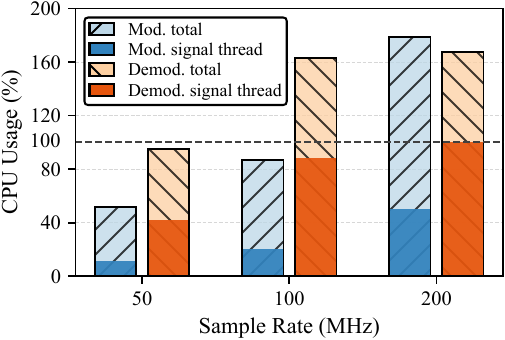}
\caption{Host-side total and signal processing thread CPU usage under proportional sample-rate, bandwidth, FFT-size, and CP-length scaling.}
\label{fig:realtime_cpu}
\end{figure}

Fig.~\ref{fig:realtime_cpu} shows that both the total CPU load and the signal processing thread CPU load increase markedly with sample rate for the transmitter and the receiver. The reported CPU percentage is normalized to a single core and can therefore exceed $100\%$. For the modulator, the CPU load of the signal-processing thread increases from 11.3\% at 50~MHz to 50.4\% at 200~MHz. In contrast, the demodulator exhibits a steeper rise in CPU load, climbing from 42.3\% to 100.5\%, indicating that the receive-side signal processing path is the dominant bottleneck and determines the practical real-time boundary of the current host-based implementation.

\begin{table}[!htbp]
\caption{Representative host-side processing time and normalized processing load for the modulator and demodulator.}
\label{tab:realtime_load}
\centering
\footnotesize
\renewcommand{\arraystretch}{1.1}
\setlength{\tabcolsep}{2pt}
\resizebox{\columnwidth}{!}{%
\begin{tabular}{lccccc}
\toprule
\makecell{\textbf{Sample}\\\textbf{Rate}} & \makecell{\textbf{Mod.}\\\textbf{Time (ms)}} & \makecell{\textbf{Mod.}\\\textbf{Proc. Load}} & \makecell{\textbf{Demod.}\\\textbf{Time (ms)}} & \makecell{\textbf{Demod.}\\\textbf{Proc. Load}} & \makecell{\textbf{RX Queue}\\\textbf{Drops}} \\
\midrule
50~MHz  & 0.21 & 8.9\%  & 0.81 & 35.3\%  & 0  \\
100~MHz & 0.42 & 18.1\% & 1.95 & 84.7\%  & 0  \\
200~MHz & 1.10 & 47.8\% & 3.95 & 171.5\% & $>1600$ \\
\bottomrule
\end{tabular}%
}
\end{table}

In Table~\ref{tab:realtime_load}, the processing load is defined as frame processing time divided by frame duration. The results show that the proportionally scaled 50~MHz and 100~MHz configurations are stable operating points, whereas the 200~MHz/4096-FFT configuration exceeds the real-time capacity of the current host-only demodulator path. In this case, the demodulator processing load rises above 100\% and the RX frame queue reports more than 1600 dropped frames, indicating persistent backlog buildup and practical frame drops.

\begin{table}[!htbp]
\caption{Monostatic sensing CPU load and processing time under different strides and processing modes.}
\label{tab:sensing_runtime}
\centering
\footnotesize
\renewcommand{\arraystretch}{1.1}
\setlength{\tabcolsep}{2pt}
\resizebox{\columnwidth}{!}{%
\begin{tabular}{ccccc}
\toprule
\multirow{2}{*}{\makecell{\textbf{Stride}\\${M_D}$}} & \multicolumn{2}{c}{\textbf{FFT On}} & \multicolumn{2}{c}{\textbf{FFT Off}} \\
\cmidrule(lr){2-3}\cmidrule(lr){4-5}
 & \textbf{MTI Off} & \textbf{MTI On} & \textbf{MTI Off} & \textbf{MTI On} \\
\midrule
1  & 100.0\% / 1.21~ms & 100.0\% / 2.54~ms & 33.7\% / 0.19~ms & 100.0\% / 1.53~ms \\
2  & 62.9\% / 1.21~ms & 100.0\% / 2.58~ms & 20.8\% / 0.20~ms & 76.7\% / 1.54~ms \\
5  & 28.7\% / 1.23~ms & 50.9\% / 2.52~ms & 12.0\% / 0.22~ms & 34.2\% / 1.54~ms \\
10 & 17.5\% / 1.26~ms & 28.7\% / 2.56~ms & 9.2\% / 0.28~ms & 20.4\% / 1.56~ms \\
20 & 11.7\% / 1.37~ms & 17.3\% / 2.48~ms & 7.4\% / 0.39~ms & 13.0\% / 1.59~ms \\
\bottomrule
\end{tabular}%
}
\vspace*{-2pt}
\end{table}

Table~\ref{tab:sensing_runtime} reports the average sensing-thread CPU load and processing time at 100~MHz and FFT size $1024$ for four processing modes. The low-stride cases reveal the real-time boundary: $M_D=1$ causes dropped frames in three of the four modes, and the full-processing case with FFT and MTI enabled still exhibits drops at $M_D=2$. By contrast, all tested cases with $M_D \ge 5$ complete without recorded drop events. Increasing $M_D$ therefore provides a practical tradeoff between slow-time resolution and real-time margin, while the per-batch processing time changes much less because each sensing batch still contains the same number of sampled OFDM symbols.

\subsubsection{Communication Performance}

\begin{figure}[!htbp]
\centering
\includegraphics[width=0.35\textwidth]{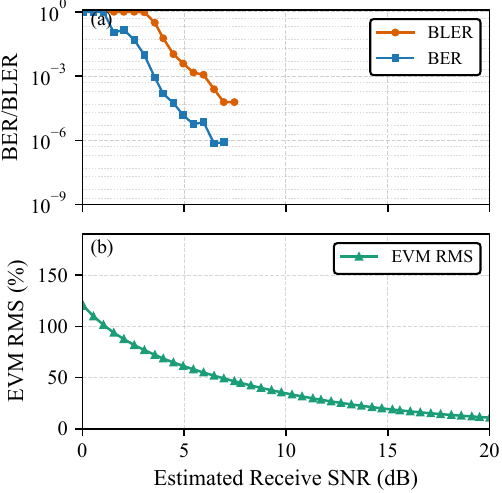}
\caption{BER, BLER, and EVM versus estimated receive SNR.}
\label{fig:comm_perf}
\end{figure}

We further evaluate the communication performance of OpenISAC by transmitting 16384 packets of 1024 bytes at each operating point under the same 50~MHz X310-to-B210 OFDM configuration used in the sensing experiments. As shown in Fig.~\ref{fig:comm_perf}(a), both BLER and BER remain zero down to about 7.8~dB, after which the BLER increases as the estimated receive SNR decreases and reaches 1 near 2.0~dB. Fig.~\ref{fig:comm_perf}(b) shows the corresponding modulation-quality degradation, where the average EVM RMS increases from 10.8\% at high SNR and exceeds 100\% below about 1.0~dB.

\subsubsection{Monostatic Sensing}

\begin{figure}[!htbp]
\centering
\subfloat[Without MTI]{
\includegraphics[width=0.47\linewidth]{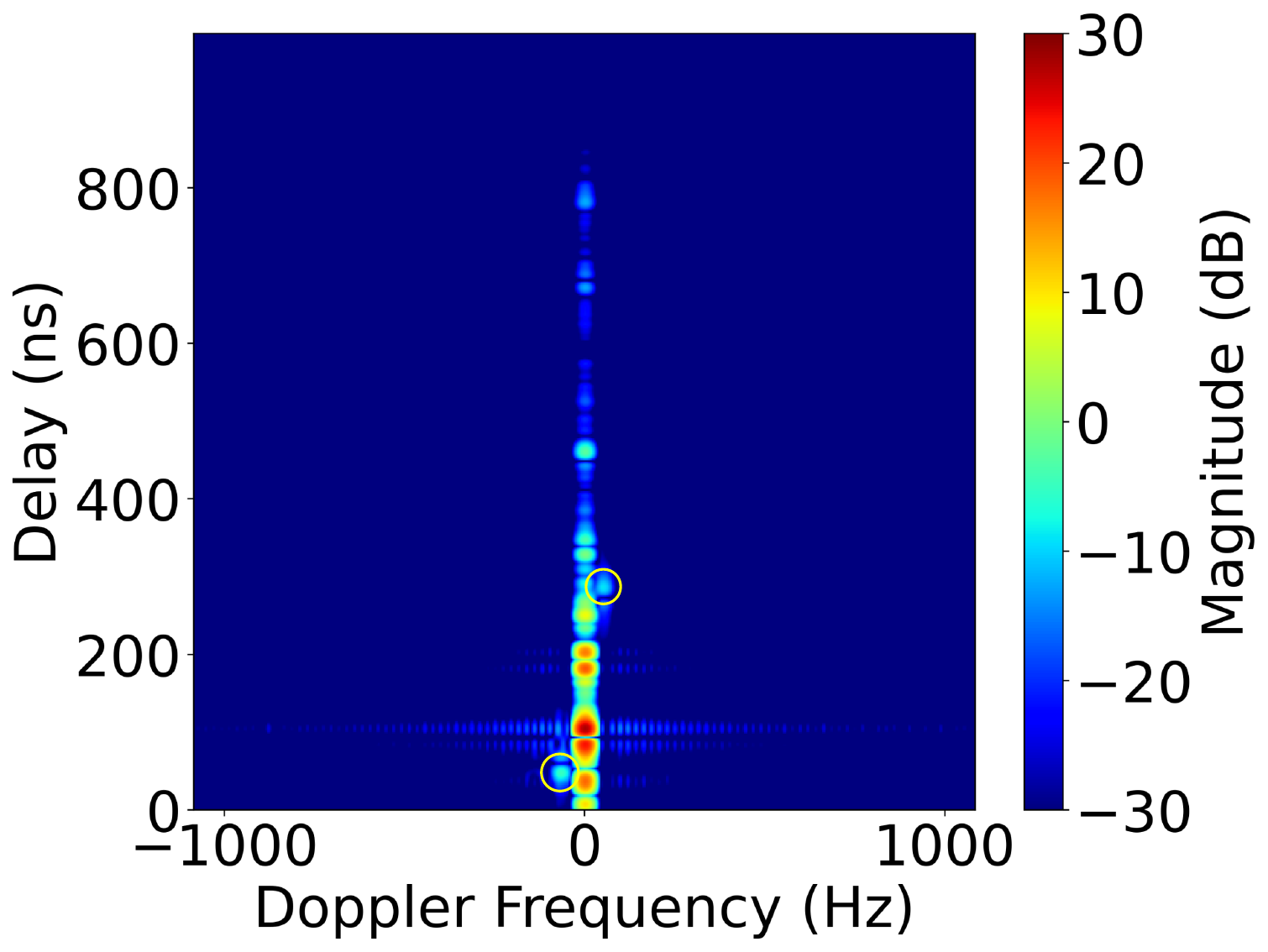}
\label{WithoutMTI}
}
\subfloat[With MTI]{
\includegraphics[width=0.47\linewidth]{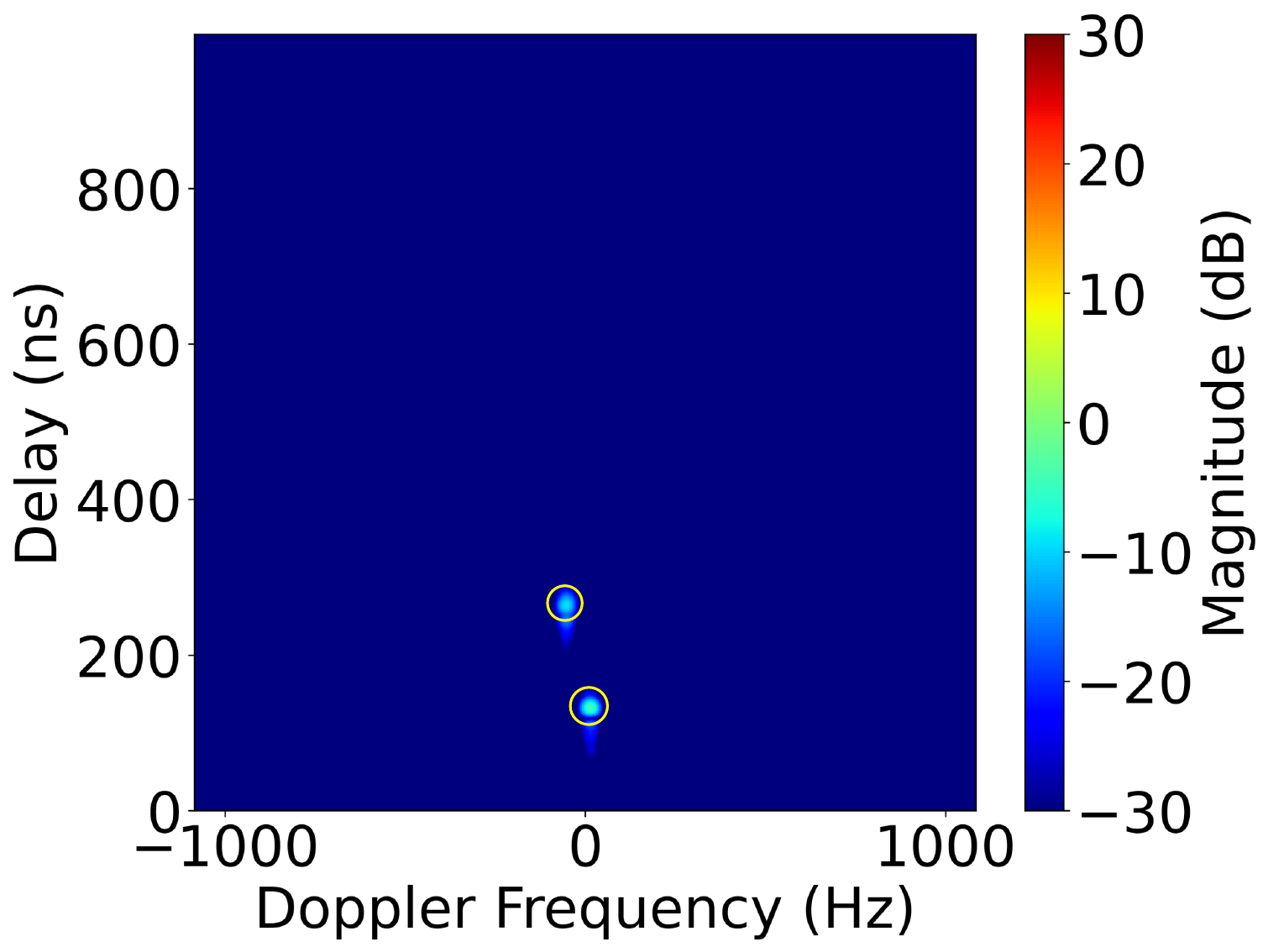}
\label{WithMTI}
}

\caption{Monostatic sensing results in a two-target dynamic environment.}
\label{fig:monostatic_sensing}
\end{figure}
Fig.~\ref{fig:monostatic_sensing} illustrates the monostatic sensing performance in a two-target environment. 
In Fig.~\ref{WithoutMTI}, two targets have delay-Doppler pairs $(51~\mathrm{Hz}, 285~\mathrm{ns})$ and $(-65~\mathrm{Hz}, 45~\mathrm{ns})$ respectively with MTI processing off. The delay-Doppler map is dominated by strong static clutter and self-interference around zero Doppler. The targets are relatively weak compared with the clutter, which could cause misdetections.
After applying MTI clutter suppression, Fig.~\ref{WithMTI} shows that the static components are effectively removed, resulting in a much cleaner background and clearly demonstrating the clutter suppression capability of the proposed MTI processing. The two targets with delay-Doppler pairs $(-58~\mathrm{Hz}, 265~\mathrm{ns})$ and $(10~\mathrm{Hz}, 131~\mathrm{ns})$ are more clear. Note that the delay-Doppler spectra are captured from the real-time system, and the two delay-Doppler pairs in Fig.~\ref{WithoutMTI} and Fig.~\ref{WithMTI} are not identical, since it is challenging to keep the delays and Dopplers of both moving targets exactly the same across repeated measurements.

\begin{figure}[!htbp]
\centering
\includegraphics[width=0.3\textwidth]{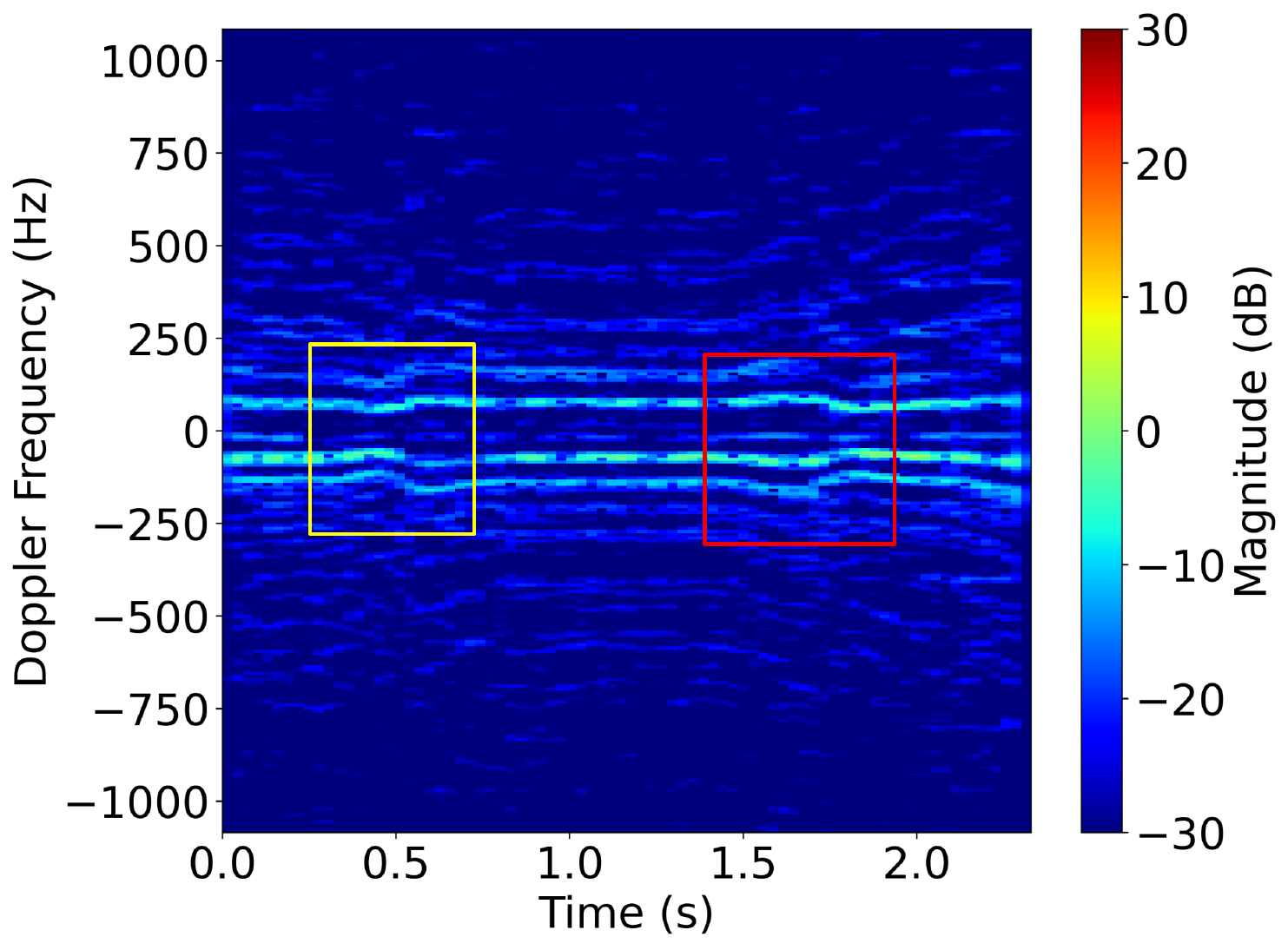}
\caption{Micro-Doppler spectrum of a Mavic Air 3S drone.}
\label{fig:micro-Doppler_Mono}
\end{figure}
Fig.~\ref{fig:micro-Doppler_Mono} illustrates the monostatic micro-Doppler spectrogram of a hovering Mavic Air 3S. The zero-Doppler return represents quasi-static scattering from the UAV body. Symmetric, equidistant ridges correspond to the rotating rotor blades. The spacing between these ridges reflects the rotor angular velocity, while the overall Doppler spread indicates the maximum radial velocity of the blade tips. The yellow box highlights a descent maneuver, where a reduction and subsequent restoration of lift leads to blade deceleration and re-acceleration, visible as a contraction-expansion of the micro-Doppler ridges. Conversely, the red box marks an ascent maneuver, where the increase and subsequent decrease in rotor speed produces an opposing expansion-contraction pattern.

\subsubsection{Bistatic Sensing}

\begin{figure}[!htbp]
\centering
\subfloat[Without OTA synchronization]{
\includegraphics[width=0.47\linewidth]{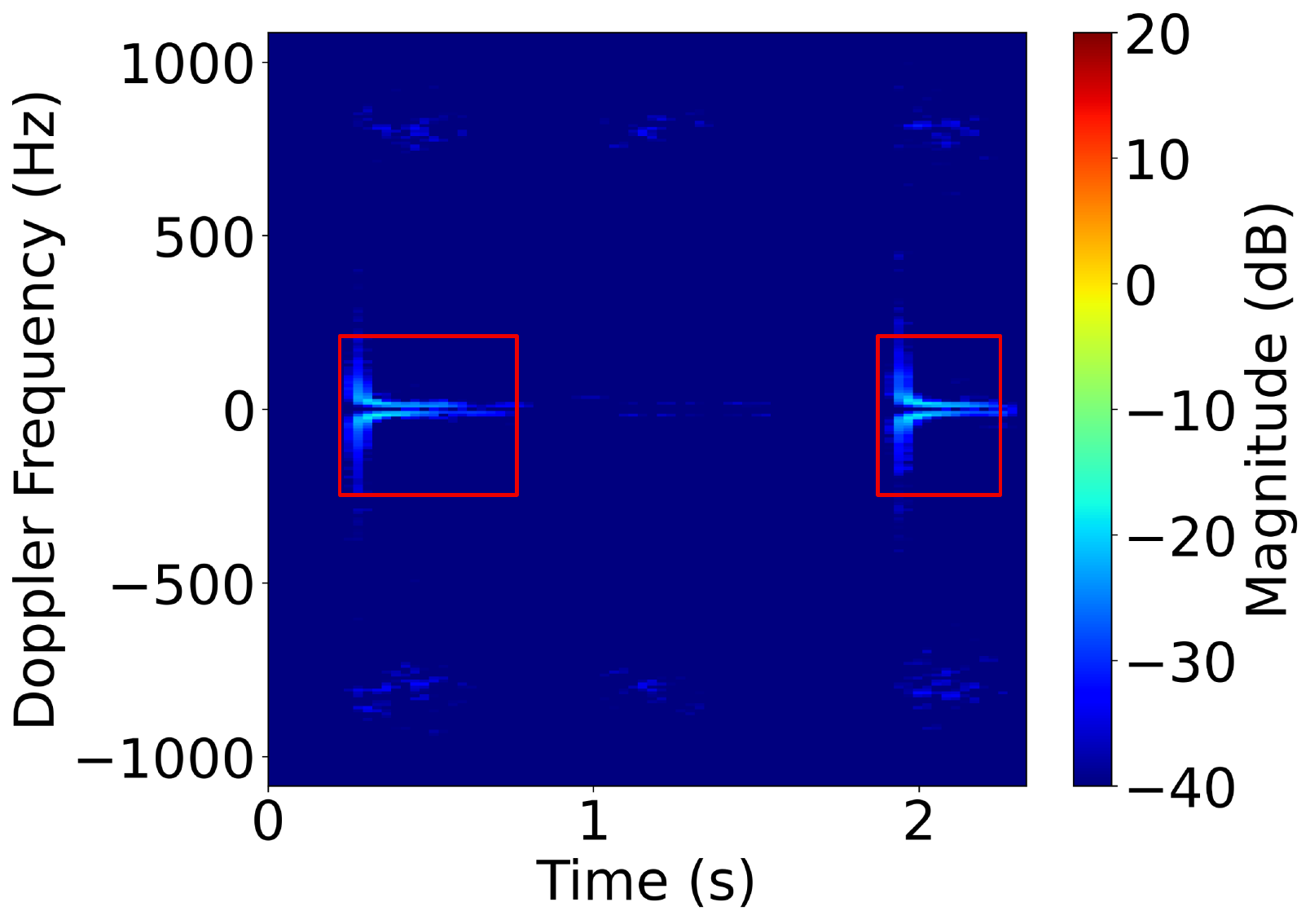}
\label{Without_Synchronization}
}
\subfloat[With OTA synchronization]{
\includegraphics[width=0.47\linewidth]{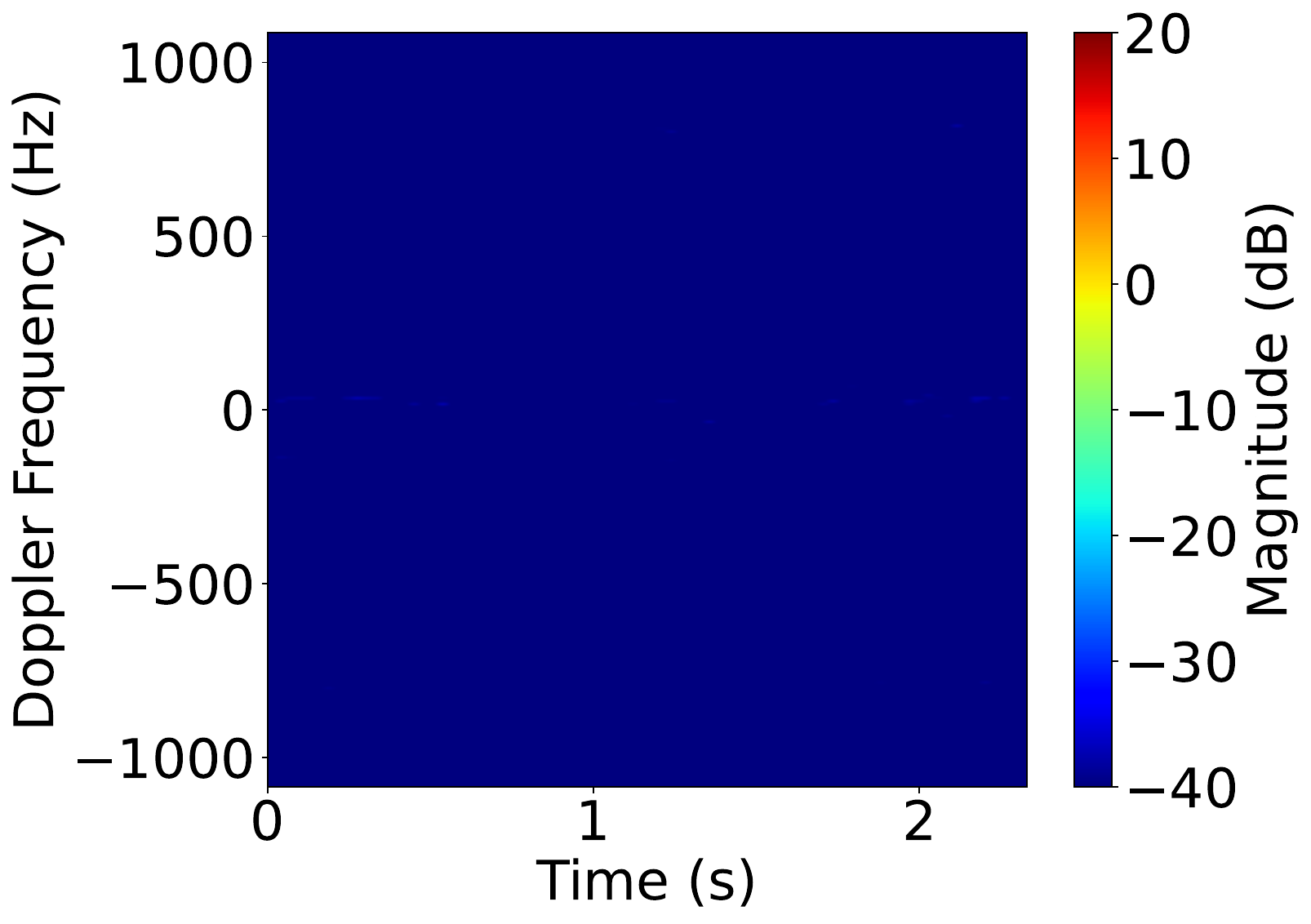}
\label{With_Synchronization}
}
\caption{Bistatic micro-Doppler Spectrum of a static environment.}
\label{fig:bistatic_sensing}
\end{figure}

Fig.~\ref{fig:bistatic_sensing} compares bistatic micro-Doppler results in a static environment. Without OTA synchronization (Fig.~\ref{Without_Synchronization}), uncompensated SIO and conventional communication-only timing corrections lead to piecewise-linear drifts and integer-sample jumps, as explained in Sec.~\ref{subsec:UE_Sync}. When micro-Doppler processing is performed at a fixed delay bin, these abrupt delay jumps manifest as discontinuities in the micro-Doppler spectrum, as indicated by the red boxes in Fig.~\ref{Without_Synchronization}.

With OTA synchronization enabled, the UE maintains a continuous sensing timing estimate $\hat{k}^{\mathrm{sens}}_{\tau,\gamma}$ and compensates the sensing channel symbols according to \eqref{eq:bistatic_sensing_symbols_comp}. This effectively removes the SIO-induced delay drift and suppresses the integer-jump artifacts, yielding a stable slow-time stream suitable for long coherent processing. Consequently, as shown in Fig.~\ref{With_Synchronization}, the spectral discontinuities are eliminated, resulting in a clean micro-Doppler spectrum free from the spurious components observed in Fig.~\ref{Without_Synchronization}.

To further quantify synchronization stability and its impact on clutter suppression, we evaluate the MTI suppression ratio (MSR), defined as the ratio of the signal energy before and after MTI filtering:
\begin{equation}
\label{eq:MSR_def}
\mathrm{MSR} = \frac{\sum_{m=m_{\mathrm{start}}}^{m_{\mathrm{start}}+M_{\mathrm{avg}}}{\sum_{n=0}^{N-1}{\left| (\grave{\boldsymbol{F}}_{\mathrm{UE}})_{n,m} \right|^2}}}{\sum_{m=m_{\mathrm{start}}}^{m_{\mathrm{start}}+M_{\mathrm{avg}}}{\sum_{n=0}^{N-1}{\left| (\tilde{\boldsymbol{F}}_{\mathrm{UE}})_{n,m} \right|^2}}},
\end{equation}
where $(\grave{\boldsymbol{F}}_{\mathrm{UE}})_{n,m}$ and $(\tilde{\boldsymbol{F}}_{\mathrm{UE}})_{n,m}$ denote the bistatic TF-domain samples before and after MTI processing, respectively. In our experiment, the evaluation window spans all subcarriers, and the slow-time averaging length is set to $M_{\mathrm{avg}}=10{,}000$.

MSR is used here as a synchronization-stability proxy in a static scene. To see its connection with synchronization error, consider the single-static-component case of \eqref{eq:ue_rx_channel}, where we keep only one path and set $f_{D,1}=0$. Omitting the residual frequency term for clarity, the pre-MTI TF-domain sample can be written as
\begin{equation}
(\grave{\boldsymbol{F}}_{\mathrm{UE}})_{n,m} \approx \alpha _1 e^{-j2\pi n\Delta f \left( \tau _1+\bar{\tau}_{d,\gamma,mN_s} \right)}.
\label{eq:msr_clutter_model}
\end{equation}
For intuition, if a two-pulse MTI is used, then
\begin{equation}
\begin{aligned}
(\tilde{\boldsymbol{F}}_{\mathrm{UE}})_{n,m}
&\approx \alpha _1 e^{-j2\pi n\Delta f\tau _1}
\\
&\quad \cdot \left(e^{-j2\pi n\Delta f\bar{\tau}_{d,\gamma,mN_s}}-e^{-j2\pi n\Delta f\bar{\tau}_{d,\gamma,(m-1)N_s}}\right).
\end{aligned}
\label{eq:msr_sync_relation}
\end{equation}
Taking the squared magnitude of \eqref{eq:msr_sync_relation}, the common phase term is canceled and we obtain
\begin{equation}
\left|(\tilde{\boldsymbol{F}}_{\mathrm{UE}})_{n,m}\right|^2
=4|\alpha_1|^2\sin^2\!\bigl(\pi n\Delta f(\bar{\tau}_{d,\gamma,mN_s}\!-\!\bar{\tau}_{d,\gamma,(m-1)N_s})\bigr).
\label{eq:msr_sync_relation_power}
\end{equation}
Equation \eqref{eq:msr_sync_relation_power} shows that larger inter-frame timing fluctuations increase the post-MTI residual energy and thus reduce the achievable MSR. Although our implementation uses an IIR high-pass MTI filter rather than a two-pulse canceller, the same trend remains.

\begin{figure}[!htbp]
\centering
\includegraphics[width=0.40\textwidth]{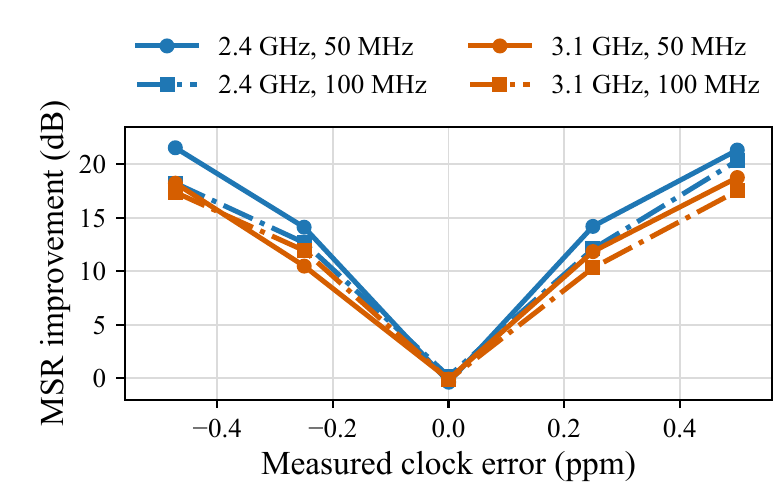}
\caption{MSR improvement of the proposed OTA synchronization under different center frequencies and bandwidths.}
\label{fig:msr_sweep}
\end{figure}

To validate the performance of OTA synchronization under different RF configurations, we measure the MSR at two center frequencies, $f_c\in\{2.4,3.1\}$~GHz, and two bandwidths, $B\in\{50,100\}$~MHz, in a static scene. The UE reference OCXO is driven by a USB-controlled DAC, which allows us to intentionally vary the reference clock error from approximately $-0.47$ to $0.50$~ppm and evaluate the corresponding MSR improvement. As shown in Fig.~\ref{fig:msr_sweep}, the improvement is close to 0~dB near zero clock error, about 10--14~dB around $\pm0.25$~ppm, and about 17--22~dB around $\pm0.5$~ppm. The consistent improvement across the tested center frequencies and bandwidths demonstrates the effectiveness of the proposed OTA synchronization.

\vspace*{-16pt}
\begin{figure}[!htbp]
\centering
\subfloat[Micro-Doppler Spectrum]{
\includegraphics[width=0.48\linewidth]{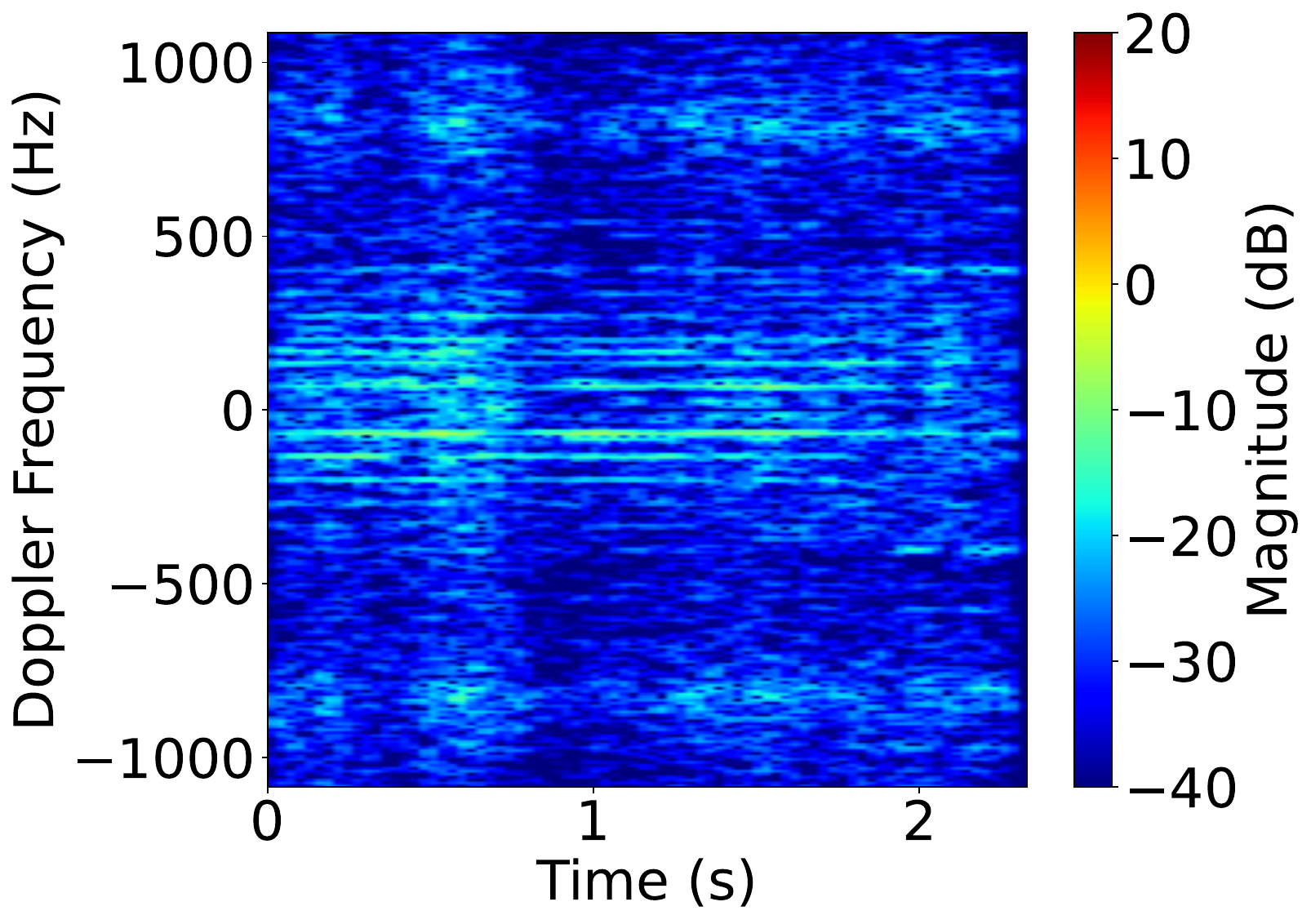}
\label{bistatic_md}
}
\subfloat[Delay-Doppler Spectrum]{
\includegraphics[width=0.45\linewidth]{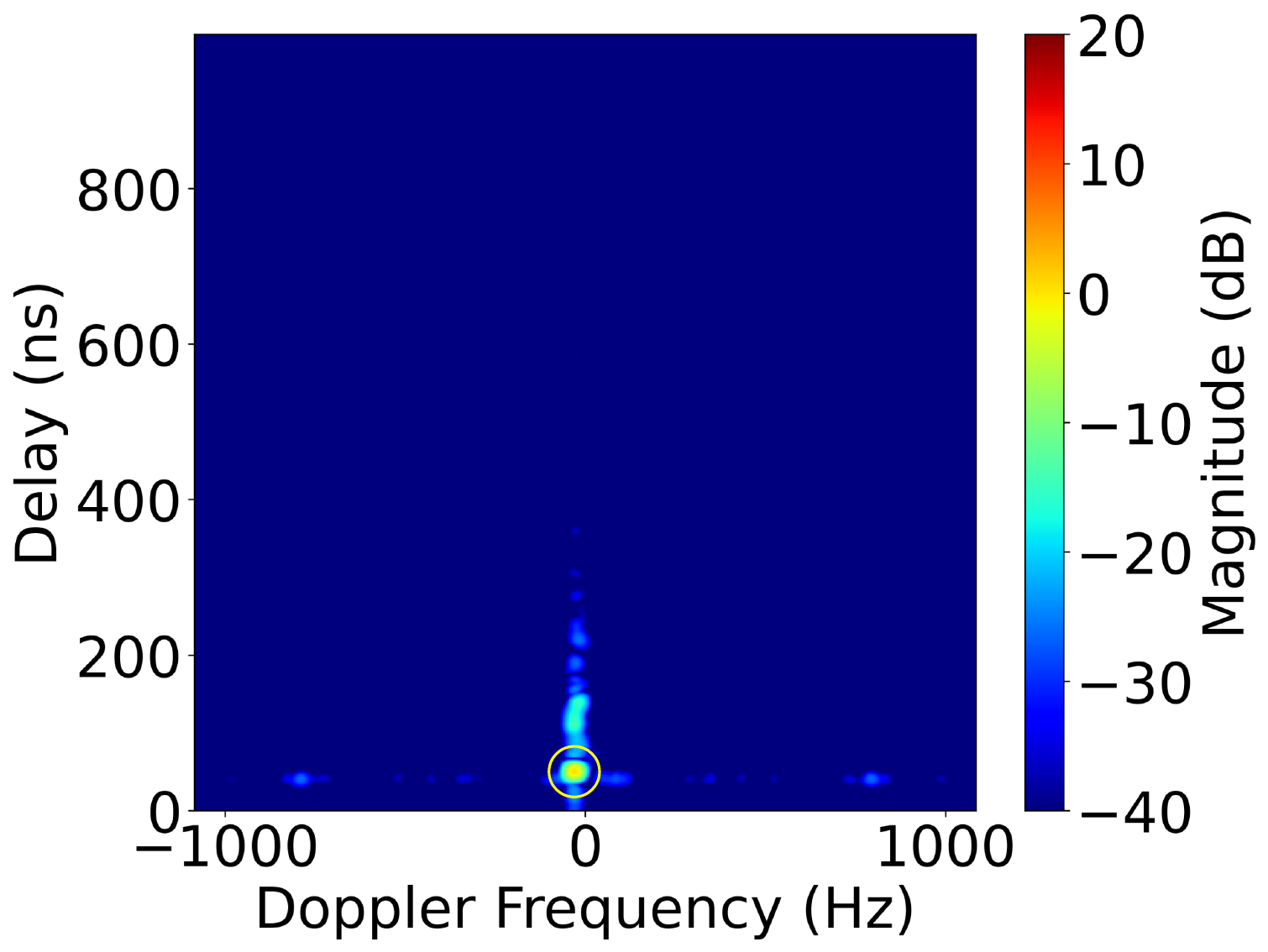}
\label{bistatic_dd}
}
\caption{Bistatic sensing results of a UAV with OTA synchronization.}
\label{fig:bistatic_uav}
\end{figure}

We further validate the bistatic sensing performance using a Mavic Air 3S as the target. Fig.~\ref{bistatic_md} displays the micro-Doppler spectrum of a hovering UAV, clearly exhibiting the characteristic symmetric blade ridges consistent with monostatic observations (Fig.~\ref{fig:micro-Doppler_Mono}). For the delay-Doppler spectrum in Fig.~\ref{bistatic_dd}, the UAV was moved slowly to ensure visibility after MTI filtering. The resulting map reveals a distinct target return at $(-32.3~\mathrm{Hz}, 49.7~\mathrm{ns})$, demonstrating that the proposed OTA synchronization enables practical real-time bistatic sensing.

\section{Conclusion} \label{conclusion}
This paper presents OpenISAC, a fully open-source, real-time OFDM-ISAC platform that supports both monostatic and bistatic sensing on USRP hardware using a hybrid C++/Python architecture. Its host-based real-time implementation provides an accessible and practical platform for OFDM-ISAC experimentation without relying on proprietary software or FPGA-centric development. Experimental results validate the system's capabilities in both monostatic and bistatic scenarios, demonstrating accurate delay-Doppler detection and clear micro-Doppler signature extraction. OpenISAC also includes practical support for wire-free bistatic operation through an OTA synchronization option. Future versions of OpenISAC will extend the current SISO prototype toward multi-antenna operation. We are developing multi-receive-channel monostatic sensing with multi-USRP synchronization and calibration, and future work will further support angle sensing, multi-transmit antennas, multi-UE receive antennas, and multi-stream transmission.

\bibliographystyle{IEEEtran}
\bibliography{IEEEabrv,reference}

\end{document}